\documentclass[12pt]{article}

\usepackage{epsfig,epsf}
\usepackage{graphics}
\usepackage{booktabs}
\usepackage{cite}
\usepackage{geometry}
\usepackage{xcolor}

\usepackage{amsmath}
\usepackage{amsthm}
\usepackage{amsfonts}
\usepackage{amssymb}
\usepackage{bm}

\usepackage{slashed}
\usepackage{braket}

\DeclareMathOperator{\im}{Im}

 \definecolor{darkgreen}{rgb}{0.1, 0.4, 0.}
 
  \definecolor{darkorange}{rgb}{0.9, 0.5, 0.}

\begin{document}
\allowdisplaybreaks
\thispagestyle{empty}

\begin{flushright}
{\small
IPPP/17/70\\[0.1cm]
\today
}
\end{flushright}

\vskip2cm
\begin{center}
\textbf{\Large\boldmath Reconstruction of top-quark mass effects in Higgs\\[0.1cm]
pair production and other gluon-fusion processes}
\\
\vspace{2cm}
{\sc Ramona Gr\"ober}, {\sc Andreas Maier} and {\sc Thomas Rauh}\\[0.5cm]
\vspace*{0.5cm} {\it
IPPP, Department of Physics,
University of Durham,\\
DH1 3LE, United Kingdom}

\def\thefootnote{\arabic{footnote}}
\setcounter{footnote}{0}

\vskip3cm
\textbf{Abstract}\\
\vspace{1\baselineskip}
\parbox{0.9\textwidth}{
We propose a novel method for the treatment of top-quark mass
effects in the production of $H^{(*)}$, $HH$, $HZ$ and $ZZ$ final states
in gluon fusion. We show that it is possible to reconstruct
the full top-quark mass dependence of the virtual amplitudes
from the corresponding large-$m_t$ expansion and the non-analytic
part of the amplitude near the top-quark threshold $\hat{s}=4m_t^2$
with a Pad\'e ansatz. The reliability of our method is clearly
demonstrated by a comparison with the recent NLO result for Higgs
pair production with full top-quark mass dependence.
}

\end{center}


\newpage
\setcounter{page}{1}


\section{Introduction\label{sec:intro}}

Gluons are ubiquitous at the LHC, and gluon fusion is among the
phenomenologically most interesting production mechanisms. Specifically,
the production of final states including one or more Higgs bosons is
typically dominated by gluon fusion, with a virtual top-quark loop
mediating the interaction to the Higgs bosons.

Precise predictions for such processes are indispensable for measuring
the properties of the Higgs boson. On the one hand, gluon fusion
processes experience large $K$-factors.\footnote{
See~\cite{Ahrens:2008qu,Ahrens:2008nc,Ebert:2017uel} for a discussion of
'timelike' logarithms in gluon fusion and their resummation which
reduces the size of perturbative corrections significantly.} Examples
include a $K$-factor of $2.3$ for single Higgs and $1.7$ for Higgs pair
production at next-to-leading order
(NLO)~\cite{Spira:1995rr,Harlander:2005rq,Anastasiou:2006hc,Aglietti:2006tp,Anastasiou:2016cez,Borowka:2016ehy,Borowka:2016ypz}
which clearly demonstrates the importance of taking higher-order
corrections into account. On the other hand, calculating these
higher-order corrections is extremely challenging. Gluon fusion is a
loop-induced process, and the top-quark mass introduces an additional
scale in the loop integrals. While the NLO
corrections to single-Higgs production have been known analytically for
some time~\cite{Spira:1995rr,Harlander:2005rq,Anastasiou:2006hc,Aglietti:2006tp}, the calculation of
NLO corrections to processes with more than one final-state particle is
still subject of on-going work. For di-Higgs production, which requires
the evaluation of two-loop integrals with four scales, numerical results
have only become available
recently~\cite{Borowka:2016ehy,Borowka:2016ypz}.

To make higher-order computations feasible an effective field theory
(EFT), where the top quark has been integrated out in the limit
of an infinite top-quark mass, $m_t \to \infty$, has been used extensively in the literature. In this
approximation, results are available at NNNLO for single Higgs
production~\cite{Anastasiou:2015ema,Anastasiou:2016cez} and at NNLO for
Higgs pair production~\cite{deFlorian:2013jea,Grigo:2014jma}, and for
other gluon fusion processes, i.e.~$gg \to ZZ$, $gg\to H j$ at 
NNLO~\cite{Boughezal:2013uia,Chen:2014gva,Boughezal:2015dra,Boughezal:2015aha} 
and $gg\to HZ$. Beyond the infinite top mass limit, several results have 
also been obtained in the large-$m_t$ expansion (LME) for a number of 
processes listed here:
\begin{itemize}
 \item $gg\to H$: up to $1/m_t^6$ at NNLO~\cite{Harlander:2009bw,Pak:2009bx,Harlander:2009mq,Pak:2009dg,Harlander:2009my}, including $gg\to Hg$ at NLO
 \item $gg\to HH$: up to $1/m_t^{12}$ in \cite{Grigo:2015dia} and
   $1/m_t^8$ in \cite{Degrassi:2016vss} at NLO; up to $1/m_t^4$ at NNLO\cite{Grigo:2015dia}
 \item $gg\to HZ$: up to $1/m_t^8$ \cite{Hasselhuhn:2016rqt} at NLO
 \item $gg\to ZZ$: up to $1/m_t^{12}$ in \cite{Campbell:2016ivq} and
   $1/m_t^8$ in \cite{Caola:2016trd} at NLO
\end{itemize}
The expansions can be rescaled with the exact leading
order (LO) result
\begin{equation}
  \text{d}\sigma_\text{NLO}^\text{rescaled LME}/\text{d}X =
  \frac{\text{d}\sigma_\text{NLO}^\text{LME}/\text{d}X}{\text{d}\sigma_\text{LO}^\text{LME}/\text{d}X}\,\text{d}\sigma_\text{LO}^\text{exact}/\text{d}X\,,
  \label{eq:LME_rescaled}
\end{equation}
where $\text{d}\sigma/\text{d}X$ indicates the differential cross
section with respect to some quantity $X$.
For inclusive Higgs production this yields good agreement with the exact
NLO result~\cite{Spira:1995rr,Harlander:2005rq,Anastasiou:2006hc,Aglietti:2006tp}. The
comparison with the exact Higgs pair production result has however
revealed the shortcomings of the approximation~\eqref{eq:LME_rescaled}
for this process~\cite{Borowka:2016ehy,Borowka:2016ypz}. This issue is
especially pronounced when distributions are considered.

Here, we advocate a different approach, based on conformal mapping and
the construction of Pad\'e approximations from expansions in different
kinematical regimes of the amplitude. This strategy has first been
introduced for heavy-quark current correlators
$\Pi^{(j)}(q^2/(4m_q^2))$~\cite{Broadhurst:1993mw,Fleischer:1994ef} and
applied successfully up to four-loop
order~\cite{Chetyrkin:1998ix,Kiyo:2009gb,Hoang:2008qy}. The
approximation can be improved systematically by including more
information from the various kinematic limits. In fact, the three-loop
approximation is indistinguishable from the results of an exact numeric
computation~\cite{Maier:2017ypu}. In
\cite{Fleischer:1994ef}, it has also been shown for the decay $H\to
\gamma\gamma$ that a Pad\'e reconstruction of the top mass effects from
the asymptotic expansion in a large top mass yields excellent agreement
with the full NLO decay rate. Like for heavy-quark correlators and the
$H\to \gamma\gamma$ decay rate, the amplitude for Higgs production in
gluon fusion only depends on one ratio of scales and the application of
the method is straightforward. However, the amplitudes for the remaining
processes listed above depend on 4-5 scales. Pad\'e approximations based
on the LME terms alone have been used to reconstruct the interference
contribution in $gg\to ZZ$~\cite{Campbell:2016ivq}. An attempt to 
reconstruct the $gg\to HZ$ cross section has been made 
in~\cite{Hasselhuhn:2016rqt}.\footnote{The method presented below depends 
crucially on the analytic structure of the amplitude, whereas 
\cite{Hasselhuhn:2016rqt} considers Pad\'e approximants to the differential 
cross section, which is not an analytic function of the ratio 
$\hat{s}/(4m_t^2)$ near $m_t\to\infty$. Therefore, the approach used 
in~\cite{Hasselhuhn:2016rqt} does not yield an adequate description above 
the top threshold and the improvement from employing a conformal mapping 
is marginal.} In this work, we
show how such an approximation can be improved drastically by also
taking into account expansions in other kinematic regions, using Higgs
pair production as an example.

Measuring di-Higgs production at the LHC allows to directly determine
the trilinear Higgs boson self-coupling
$\lambda_3$~\cite{Djouadi:1999rca, Dolan:2012rv, Baglio:2012np}, which
serves as a probe of the shape of the Higgs potential and is a crucial
test of the mechanism of electroweak symmetry breaking in nature. While
the couplings of the Higgs boson to the gauge bosons and
third-generation fermions have been firmly established to be Standard
Model like within 10--20\%~\cite{Khachatryan:2016vau,ATLAS:2017bic,Sirunyan:2017elk},
constraining the trilinear self-coupling is highly challenging.
With $3000\,\text{fb}^{-1}$
of data the estimated bounds are $0.2 <\lambda_3/\lambda_3^{SM} < 7.0$
(neglecting systematic uncertainties) \cite{ATLASbbbb}.
Current bounds from Higgs pair production final states limit the
trilinear Higgs self-coupling between $-8.8 <\lambda_3/\lambda_3^{SM} <
15.0$ \cite{CMSbbgaga}.  Under the assumption that only the trilinear
Higgs self-coupling is modified, bounds can be obtained from single
Higgs production through the electroweak corrections
\cite{McCullough:2013rea,Gorbahn:2016uoy,Degrassi:2016wml,Bizon:2016wgr}
or from electroweak precision observables
\cite{Degrassi:2017ucl,Kribs:2017znd}. However, the current bounds are
still above the limits from perturbativity \cite{DiLuzio:2017tfn}.

\begin{figure}[t!]
 \centering
\includegraphics[width=13cm]{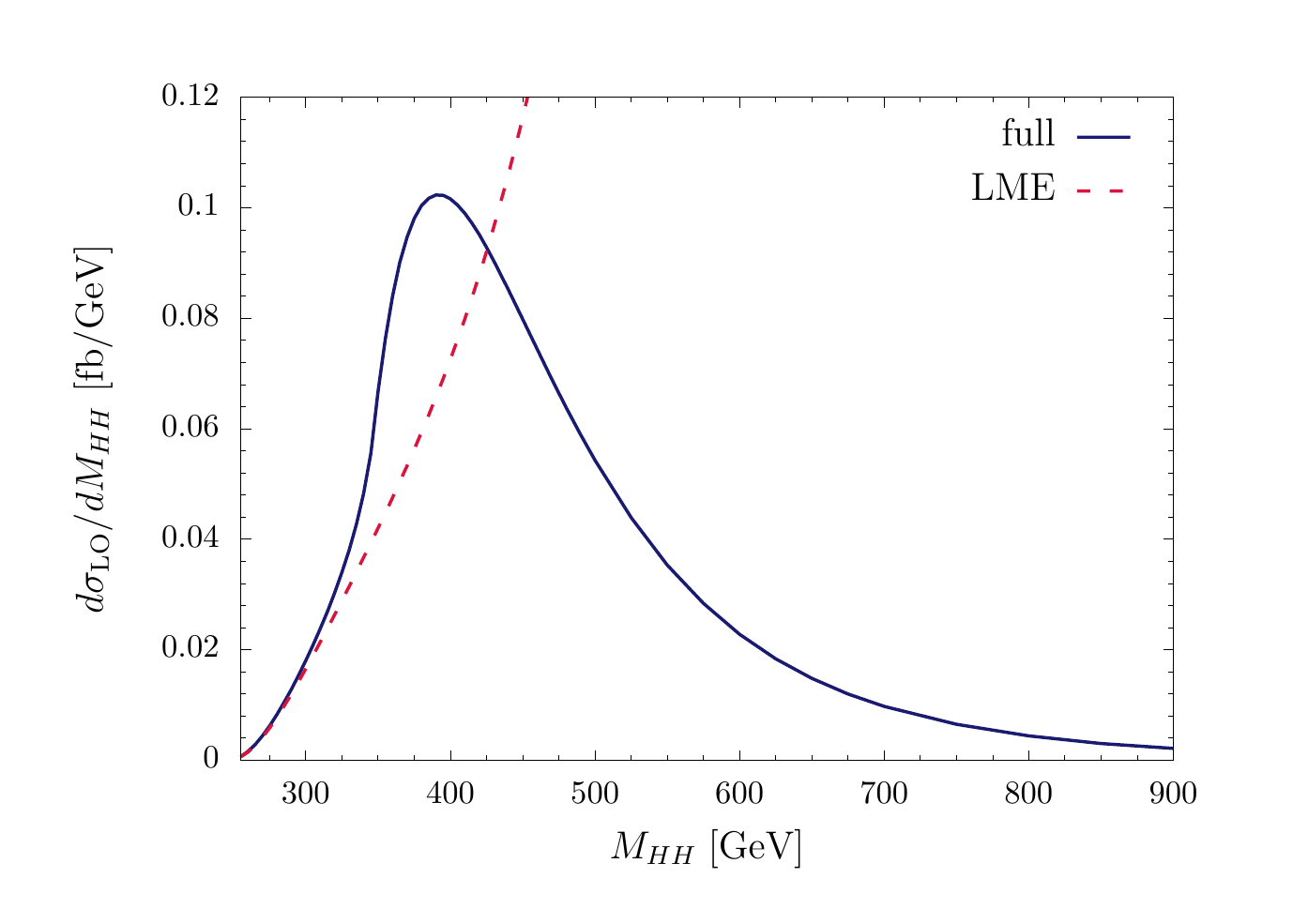}
\caption{Invariant Higgs mass distribution for the full LO cross section
  (dark blue) and the large mass expansion (LME) up to $\mathcal{O}(1/m_t^8)$
  as given in Ref.~\cite{Degrassi:2016vss} (red-dashed).\label{fig:intro}}
 \end{figure}

Precise theory predictions are crucial in the extraction of $\lambda_3$
from the cross section measurements. It is evident already at leading
order (LO) that the LME alone is not sufficient. In fact, as shown in
Figure~\ref{fig:intro}, the cross section is dominated by energies of
about 400\,GeV, whereas the LME breaks down at the top pair-production
threshold around $2m_t \approx 350\,$GeV. As we will show, constructing
Pad\'e approximations from the LME can ameliorate this problem to some
degree, but not solve it completely. The reason for this is that, above
the top threshold, the production amplitude receives non-analytic
contributions, which cannot be reproduced by the purely rational Pad\'e
approximants. Incorporating these non-analytic threshold corrections
enhances the quality of the approximation dramatically in the dominant
kinematic region and thus leads to a much improved prediction for the
total cross section.

The outline of this paper is as follows: In Section~\ref{sec:method}
we introduce our method for single Higgs production and then show
how it can be generalized to the case of Higgs pair production.
The computation of the additional input terms from the expansion
around the top threshold is described in Section~\ref{sec:threshold}.
In Section~\ref{sec:numerics} we perform a detailed comparison
of both the LO and NLO Pad\'e approximation with the full LO result and
the recent NLO results~\cite{Borowka:2016ehy,Borowka:2016ypz}, respectively.
We conclude in Section~\ref{sec:conclusion} and offer an outlook
over possible applications of our method.


\section{The method\label{sec:method}}

We first discuss the construction of a Pad\'e approximation for the simple
case of the virtual amplitude $\mathcal{A}_{gg\to H^{(*)}}$ in Section~\ref{sec:ggH_Pade}
and then generalize the approach to Higgs pair production in
Section~\ref{sec:ggHH_Pade}.

\subsection{Pad\'e approximation for $\mathbf{gg\to H^{(*)}}$\label{sec:ggH_Pade}}

\begin{figure}
 \begin{center}
    \includegraphics[width=0.8\textwidth]{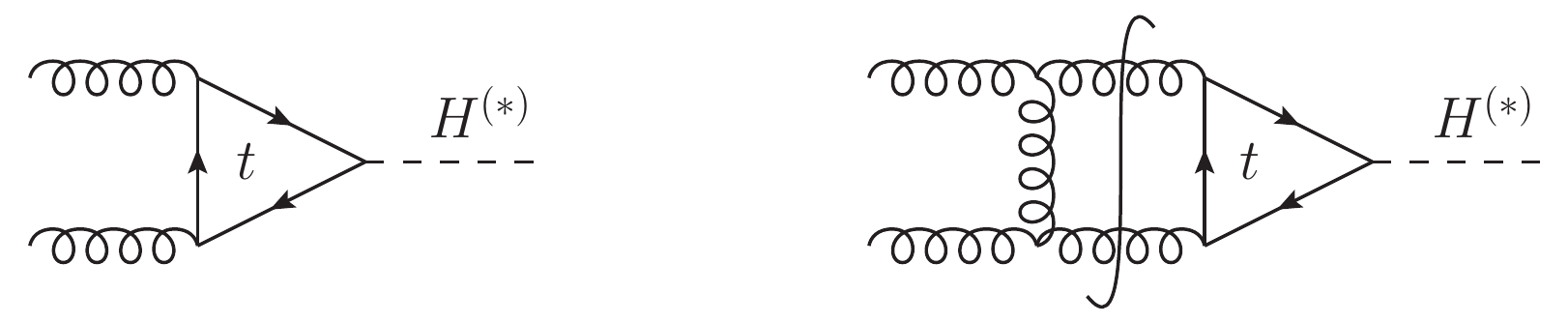}
  \caption{\label{fig:ggH}
  The LO diagram for Higgs production in gluon fusion (left)
  and an example for a NLO diagram that contains a branch
  cut starting at $\hat{s}=0$ (right).}
 \end{center}
\end{figure}

The LO diagram for the production of an off-shell
Higgs in gluon fusion is shown in Figure~\ref{fig:ggH} (left).
The corresponding amplitude can be expressed through a dimensionless
form factor $F_\triangle$ that only depends on the variable
$z=(\hat{s}+i0)/(4m_t^2)$
\begin{equation}
 \mathcal{A}^{\mu\nu}_{ab}(g(p_1,\mu,a),g(p_2,\nu,b)\to H^{(*)}(p_H)) = \frac{y_t\hat{s}}{\sqrt{2}m_t}\frac{\alpha_s}{2\pi}\delta_{ab}T_FA_1^{\mu\nu}F_\triangle(z)
 \label{eq:AggH_def}
\end{equation}
where $\hat{s}=(p_1+p_2)^2=p_H^2$, $y_t=\sqrt{2}m_t/v$ is the top Yukawa coupling,
$T_F=1/2$ and
\begin{equation}
 A_1^{\mu\nu} = g^{\mu\nu}-\frac{p_1^\nu p_2^\mu}{p_1\cdot p_2}.
 \label{eq:A1_proj}
\end{equation}
The form factor $F_\triangle$ is normalized such that
\begin{equation}
F_\triangle\xrightarrow[]{m_t\to \infty} \frac{4}{3} + {\cal O}(\alpha_s).
\end{equation}
The leading-order contribution to the form factor is analytic in the entire
complex plane with the exception of a branch cut for real $z\geq1$
due to on-shell $t\bar{t}$ cuts. At NLO, massless cuts like the
one shown in the right of Figure~\ref{fig:ggH} introduce a branch
cut starting at $z=0$. However, the branch cut can be made explicit
\begin{eqnarray}
 F_\triangle & = & F_\triangle^{1l} + \frac{\alpha_s}{\pi}\,F_{\triangle}^{2l} + \mathcal{O}(\alpha_s^2)\nonumber\\
 & = & F_\triangle^{1l} + \frac{\alpha_s}{\pi}\left[C_F F_{\triangle,C_F}^{2l} + C_A \left(F_{\triangle,C_A}^{2l}+F_{\triangle,C_A,\ln}^{2l}\ln(-4z)\right)\right] + \mathcal{O}(\alpha_s^2),
 \label{eq:Ftri_splitting}
\end{eqnarray}
such that all the $F_{\triangle,x}^{il}$ (with $i=1,2$ and $x=C_F, C_A, (C_A,\ln))$ on the right-hand side
are again analytic except for real $z\geq1$. In $F_{\triangle,C_A}^{2l}$, IR divergences in the amplitude
have been subtracted as described in Ref.~\cite{Degrassi:2016vss}. We can now apply the
conformal transformation~\cite{Fleischer:1994ef}
\begin{equation}
 z = \frac{4\omega}{(1+\omega)^2}
 \label{eq:conf_map}
\end{equation}
to map the entire complex $z$ plane onto the unit disc $|\omega|\leq1$
while the branch cut at $z\geq1$ is mapped onto the perimeter. The
physical branch $\im(z) > 0$ corresponds to the upper semicircle,
starting at $\omega(z=1) = 1$ and ending at $\omega(z\to \infty + i0) = -1$.
With this mapping, the $F_{\triangle,x}^{il}$ are analytic functions of $\omega$
inside the unit circle. We approximate them using a Pad\'e ansatz
\begin{equation}
  [n/m](\omega) = \frac{\sum\limits_{i=0}^n a_i \omega^i}{1 + \sum\limits_{j=1}^m b_j \omega^j}
  \label{eq:Pade_ansatz}
\end{equation}
with a total of $n+m+1$ coefficients. They can be fixed by imposing
conditions stemming from known expansions of the approximated
function. In many cases it is found that diagonal Pad\'e approximants
with $n=m$ provide the best description. Indeed, we find that this also
holds for our analysis. We therefore discard approximants that are too far
away from the diagonal, as detailed below.

The LME for the form factor $F_\triangle$ has been given up to terms
of the order $z^4$ in~\cite{Aglietti:2006tp}. The conformal
mapping~\eqref{eq:conf_map} transforms this into constraints on the
derivatives of the Pad\'e approximant at $\omega=0$. Furthermore the form factor
vanishes for $z\to\infty$ as $F_\triangle(z)=\mathcal{O}(1/z)$ since
$\hat{s}\sim z$ has been factored out in~\eqref{eq:AggH_def}. In a
direct approach this would imply the constraint $[n/m](\omega=-1)=0$.
Instead, we construct the Pad\'e approximant for the rescaled form
factor
\begin{equation}
 [n/m](\omega) \simeq \left[1+a_R\,z(\omega)\right] F_\triangle(z(\omega)),
\end{equation}
where $a_R$ is a free parameter. This serves a double purpose.
First, it removes the spurious constraint at $\omega=-1$ which
implies that the dimensionality of the non-linear system of equations
that determines the coefficients of the Pad\'e approximant is reduced by one.
Secondly, the variation of the parameter $a_R$ allows us to test the
stability of the ansatz and to assign an uncertainty to the reconstruction.

\begin{figure}
 \begin{center}
    \includegraphics[width=0.7\textwidth]{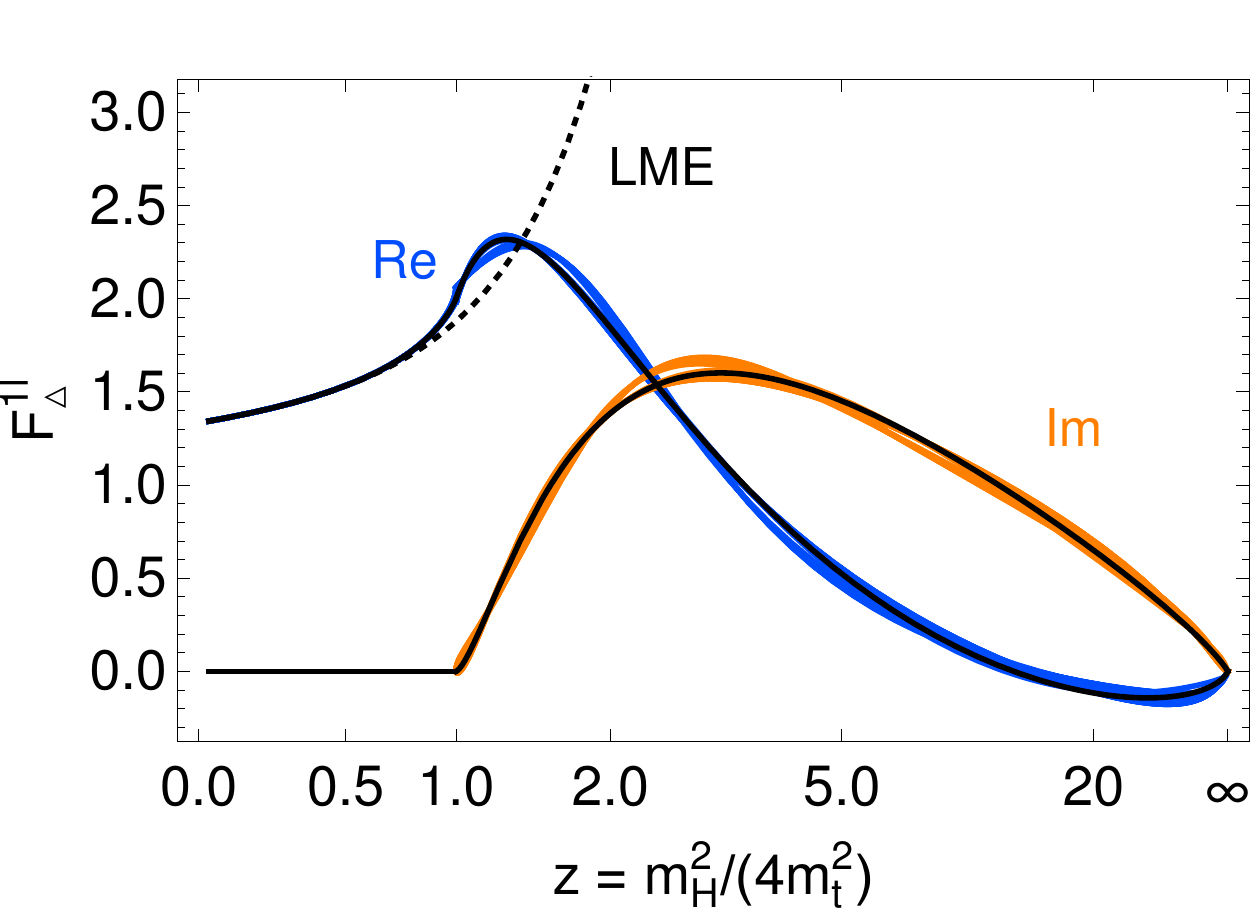}\\[0.3cm]
    \includegraphics[width=0.7\textwidth]{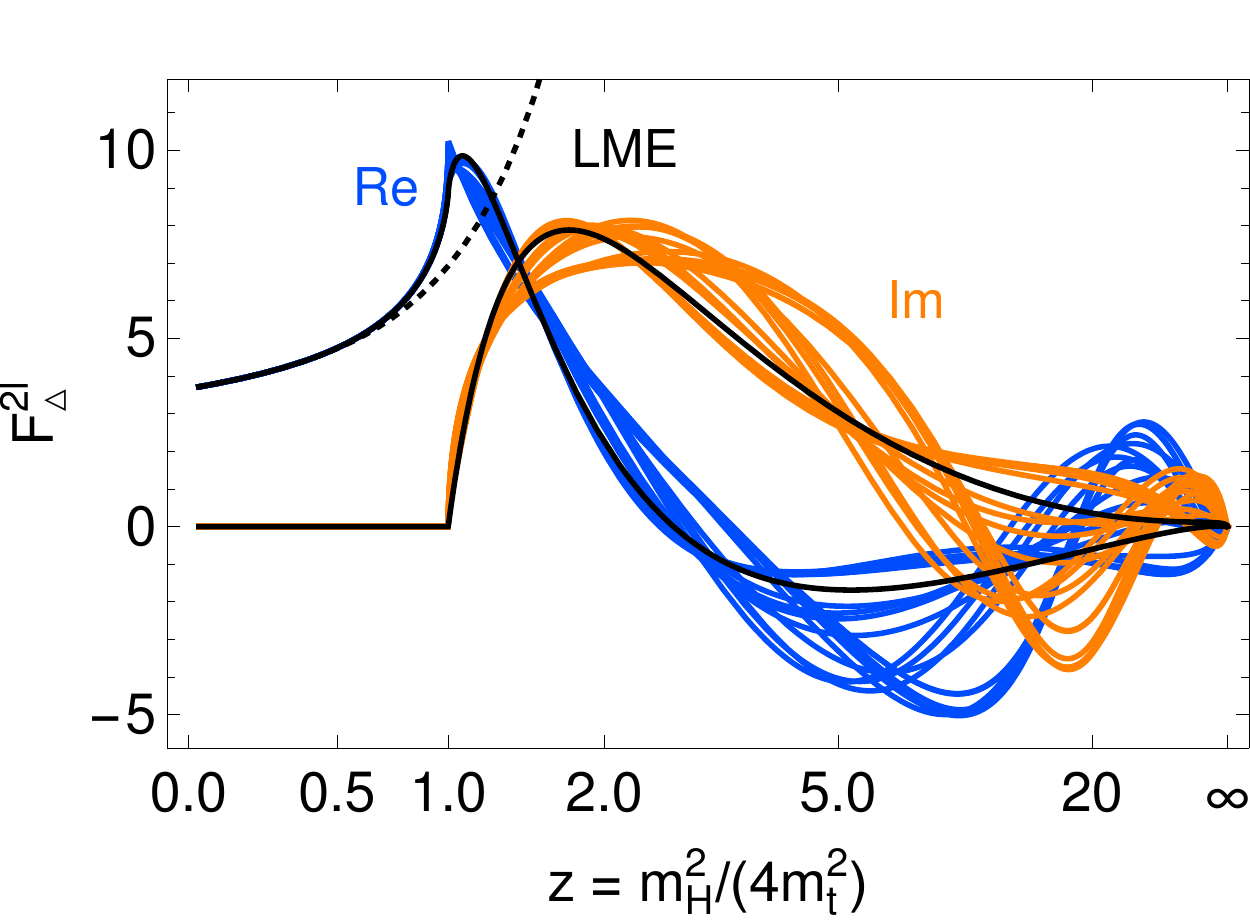}
  \caption{\label{fig:FboxLME}
  Pad\'e approximants for $F_\triangle$ at LO (top) and NLO (bottom) constructed
  using only the LME up to the order $1/m_t^8$ as input. Shown are the real/imaginary
  part of the Pad\'e approximants (blue/orange) and the exact results (black).
  We constructed in total 20 approximants of the types [1/3], [2/2] and [3/1] for
  random values of $a_R$ in the range [0.1,10], while approximants with poles in the
  rectangle $\text{Re}(z)\in[0,8]$ and $\text{Im}(z)\in[-1,1]$ have been excluded
  since they can cause unphysical resonances in the form factor.}
 \end{center}
\end{figure}

A set of Pad\'e approximants with $n+m=4$ can be constructed based only on 
the constraints from the LME up to $\mathcal{O}(z^4)$. The Pad\'e ansatz 
\eqref{eq:Pade_ansatz} has $m$ poles in the $\omega$ plane. Here, and in 
the remainder of this work, we eliminate a subset of Pad\'e approximants 
based on the positions of these poles. Since the amplitude is analytic 
inside the unit disc, the canonical selection criterion is to exclude 
approximants with poles at $|\omega|\leq1+\delta$, where $\delta>0$ 
should be chosen such that no unphysical resonances, caused by nearby 
poles, are observed in the amplitude. We find, however, that this 
criterion proves too restrictive as it excludes almost all approximants. 
Thus, we relax the selection criterion and exclude approximants with 
poles in the region corresponding to values of $z$ with 
$0\leq\text{Re}(z)\leq8$ and $-1\leq\text{Im}(z)\leq1$, thereby excluding 
poles in the vicinity of the phenomenologically relevant region 
$0.13\lesssim z\lesssim5$. We have checked the stability of the results 
under variation of the exclusion region. The result is shown in
Figure~\ref{fig:FboxLME} and compared to the exact expression for the form
factor~\cite{Spira:1995rr,Harlander:2005rq,Anastasiou:2006hc,Aglietti:2006tp}.
At LO the agreement is good, whereas at NLO the Pad\'e curves become unstable
under variations of $a_R$ and $n/m$ and show significant deviations from the
exact result for energies near and above the top threshold $z\gtrsim1$.
\begin{figure}
 \begin{center}
    \includegraphics[width=0.7\textwidth]{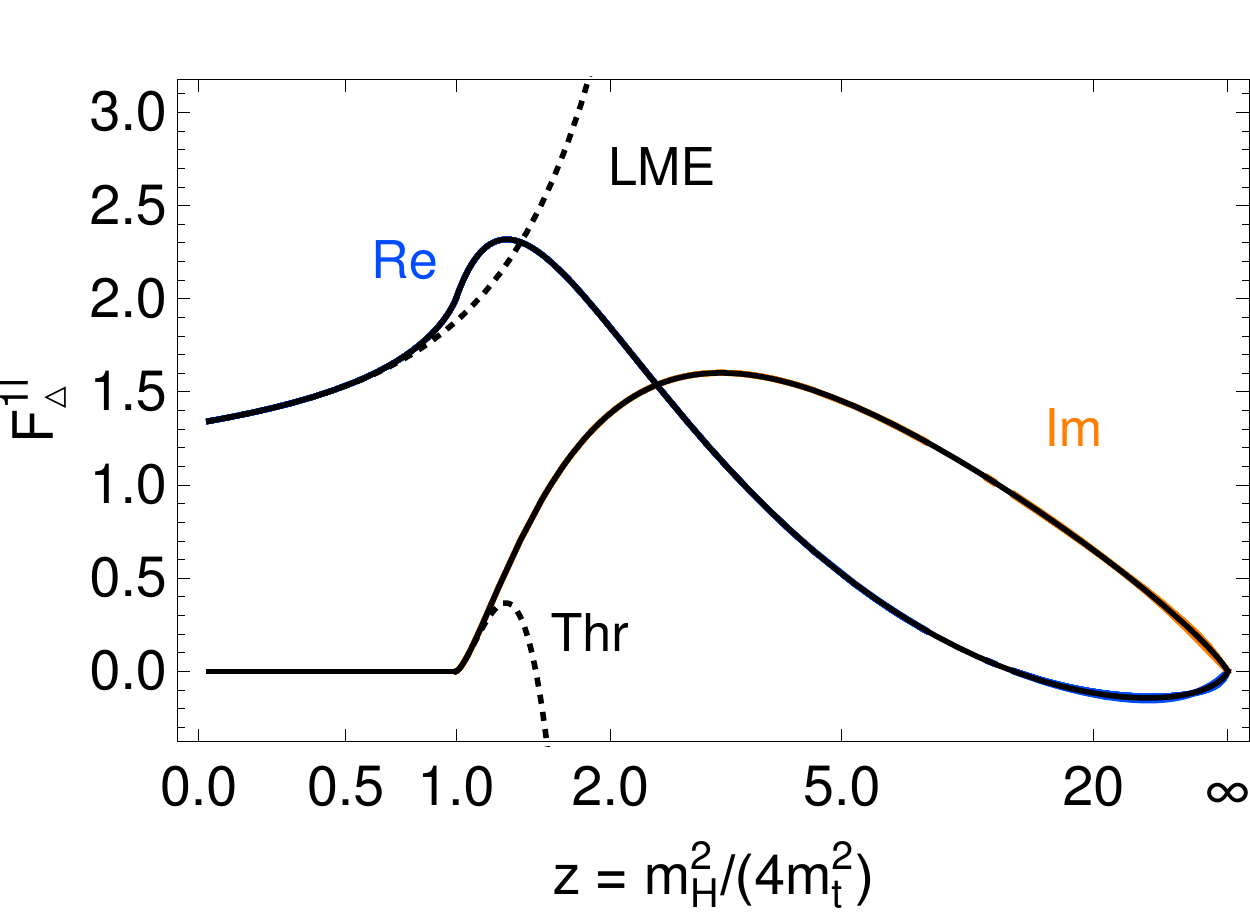}\\[1cm]
    \includegraphics[width=0.7\textwidth]{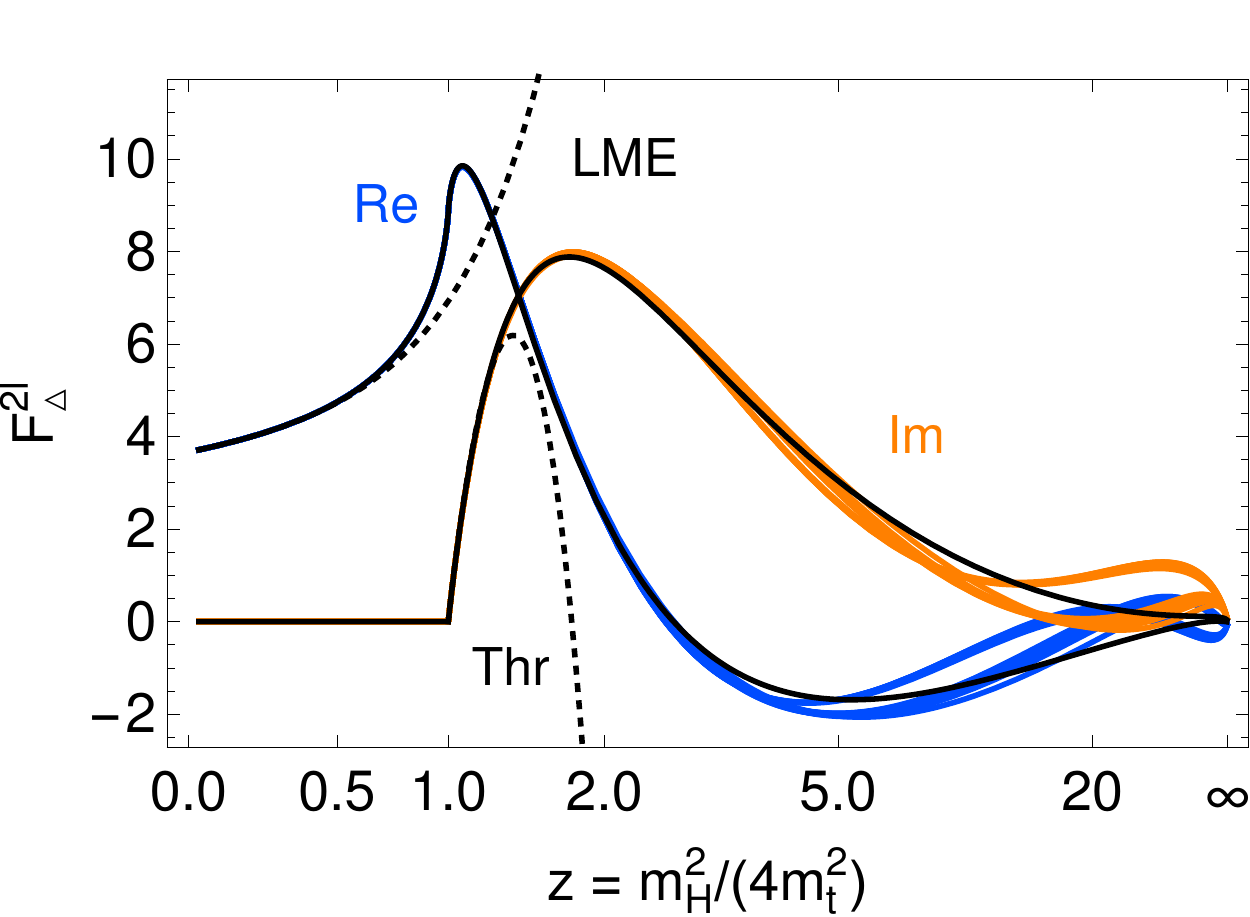}
  \caption{\label{fig:FboxLMEThr}
  We show the same comparison as in Figure~\ref{fig:FboxLME} but for Pad\'e
  approximants based on the LME and the threshold expansion. Only [5/2], [4/3],
  [3/4] and [2/5] approximants were constructed at LO and only [4/2], [3/3]
  and [2/4] approximants were constructed at NLO.}
 \end{center}
\end{figure}

We can gain some insight into this deviation by studying the expansion
of the form factor around the top threshold. In particular we are
interested in the non-analytic terms in the expansion in $(1-z)$ which
can be determined with the help of a factorization formula as discussed
below in Section~\ref{sec:threshold}. Our results take the form
\begin{eqnarray}
 F_\triangle^{1l} & \mathop{\asymp}\limits^{z\to1} & 2\pi (1-z)^{3/2} + \frac{13\pi}{3} (1-z)^{5/2} + \mathcal{O}\left((1-z)^{7/2}\right), \label{eq:triangle_threshold1}\\
 F_{\triangle,C_F}^{2l} & \mathop{\asymp}\limits^{z\to1} & \pi^2 (1-z) \ln(1-z) - \frac{\pi(40 - 3 \pi^2)}{12} (1-z)^{3/2} + \frac{2\pi^2}{3} (1-z)^2 \ln(1-z) \nonumber\\
 & &  + \mathcal{O}\left((1-z)^{5/2}\right),\label{eq:triangle_threshold2}\\
 F_{\triangle,C_A}^{2l} & \mathop{\asymp}\limits^{z\to1} & - \frac{\pi\left(3\pi ^2-4\right)}{12}(1-z)^{3/2} + \mathcal{O}\left((1-z)^{5/2}\right),\label{eq:triangle_threshold3}\\
 F_{\triangle,C_A,\ln}^{2l} & \mathop{\asymp}\limits^{z\to1} & \mathcal{O}\left((1-z)^{5/2}\right),\label{eq:triangle_threshold4}
\end{eqnarray}
where we have used the symbol $\asymp$ to denote that terms that
are analytic in $(1-z)$ have been dropped on the right-hand side.
We observe that threshold logarithms $\ln(1-z)$, which cannot be
reproduced by the Pad\'e ansatz, appear at NLO. Having determined
the coefficients of the logarithmic terms at the first two orders
we can however subtract them from the form factor and apply the
Pad\'e approximation to the subtracted function.
Taking a function $f(z)$ with the threshold expansion
\begin{equation}
 f(z) \,\,\mathop{\asymp}\limits^{z\to1}\,\, c_1 \sqrt{1-z} + c_2 (1-z)\ln(1-z) + c_3 (1-z)^{3/2} + c_4 (1-z)^2\ln(1-z) + \dots
\end{equation}
as an example we define
\begin{equation}
 \tilde{f}(z) = f(z) - c_2 s_2(z) - \left(c_4 - \frac{c_2}{3}\right) s_4(z),
\end{equation}
where $s_{2,4}$ are constructed such that their leading non-analytic
terms in the threshold expansion are given by $(1-z)\ln(1-z)$ and
$(1-z)^2\ln(1-z)$, respectively. In addition, the subtraction terms must
be analytic around $z = 0$ and at most logarithmically divergent for $z
\to \infty$.\footnote{In principle, non-logarithmic poles of the form
$z^n$ are also allowed, but these have to cancel against corresponding
poles in the Pad\'e approximation.}  Apart from these constraints, the
exact form of the subtraction functions is arbitrary.  Our choice for
the functions $s_{2,4}$ can be found in
Appendix~\ref{sec:subtractions}. The threshold expansion of $\tilde{f}$
is free of logarithms up to and including the order $(1-z)^2$. An
improved approximation of the original function $f$ is then given by
\begin{equation}
 f(z) \simeq [n/m]_{\tilde{f}}(\omega(z)) + c_2 s_2(z) + \left(c_4 - \frac{c_2}{3}\right) s_4(z),
\end{equation}
where the Pad\'e approximant $[n/m]_{\tilde{f}}$ is constructed from the expansion
terms of the subtracted function $\tilde{f}$.

In addition, the non-integer powers of $(1-z)$ in eqs.~\eqref{eq:triangle_threshold1} -- \eqref{eq:triangle_threshold4}
imply constraints on the derivatives of the Pad\'e approximation
at $\omega=1$. By using all the available constraints we can construct
approximants with a total of $n+m+1=8$ coefficients at LO and $n+m+1=7$
coefficients at NLO. The results are given in Figure~\ref{fig:FboxLMEThr}
and show perfect agreement with the exact LO form factor in the full energy
range. At NLO the agreement is excellent up to $z\sim2.5$ where tiny
deviations begin to emerge. For very large $z$, outside the phenomenologically
relevant energy range, the approximants have unphysical extrema. We suspect
that they could be removed by including information from the small $m_t$
expansion (SME) of the form factors in the construction.

An alternative implementation is obtained by performing additional
subtractions for the root terms by employing the functions $s_{1,3,5}$
in Appendix~\ref{sec:subtractions}, thereby removing all known
non-analytic terms in the expansion.
This yields the same number of constraints on the Pad\'e approximant.
In the following we will only use the subtraction functions $s_{2,4}$,
since we find no significant differences between the two approaches.

\subsection{Pad\'e approximation for $\mathbf{gg\to HH}$\label{sec:ggHH_Pade}}

The amplitude for the process $gg\to HH$ can be parametrized by two
dimensionless form factors $F_{1,2}$
\begin{equation}
 \mathcal{A}^{\mu\nu}_{ab}(g(p_1,\mu,a),g(p_2,\nu,b)\to H(p_3)H(p_4)) = y_t^2\,\frac{\alpha_s}{2\pi}\,\delta_{ab}T_F\, z\left[A_1^{\mu\nu}F_1 + A_2^{\mu\nu}F_2\right],
 \label{eq:AggHH_def}
\end{equation}
where $\hat{s}=(p_1+p_2)^2$, $\hat{t}=(p_1-p_3)^2$, $\hat{u}=(p_1-p_4)^2$,
$\hat{s}+\hat{t}+\hat{u} = 2m_H^2$, $A_1^{\mu\nu}$ is given in~\eqref{eq:A1_proj}
and
\begin{equation}
 A_2^{\mu\nu} =  g^{\mu\nu}+ \frac{ p_3^2\, p_1^{\nu}\, p^{\mu}_2
- 2\left(p_3\cdot p_2\right)p_1^{\nu}\,p_3^{\mu}
- 2\left(p_3\cdot p_1\right)p_3^{\nu}\,p_2^{\mu}
+ 2 \left(p_1\cdot p_2\right) p_3^{\mu}\,p_3^{\nu}}
{p_T^2\left(p_1\cdot p_2\right)},
 \label{eq:A2_proj}
\end{equation}
with
\begin{equation}
 p_T^2 = \frac{\hat{t}\hat{u} - m_H^4}{\hat{s}}\,.
\end{equation}
Given that there are four independent scales the dimensionless
form factors depend on three ratios
\begin{equation}
 F_i = F_i\left( r_H\equiv \frac{m_H^2}{\hat{s}}, r_{p_T}\equiv \frac{p_T^2}{\hat{s}}, z \right), \hspace{2cm}i=1,2.
 \label{eq:FiGGHH}
\end{equation}
This implies that their analytic structure is much more complicated than
it was the case for $F_\triangle$. For instance, there are branch cuts in
the complex $\hat{t}$ and $\hat{u}$ planes above the thresholds
$\hat{t}\geq4m_t^2$ and $\hat{u}\geq4m_t^2$. These are, however, not
kinematically accessible for external momenta that are both real and on shell.
Furthermore, for $z \geq 1/r_H \geq 4$ there is also a discontinuity
from cuts corresponding to the processes $gg\to t\bar{t}H$ and $H\to t\bar{t}$
which are, however, not accessible for the physical Higgs and top masses.
In the limit of small quark masses, $z\to\infty$, where this type of cut is
present, the recent analytical computation of the NLO virtual amplitudes
for Higgs plus jet production~\cite{Melnikov:2016qoc,Melnikov:2017pgf} has
revealed a rather complicated structure of logarithms in the soft and (in
particular) the collinear limit which is presently not fully understood.

Here, we take a practitioners approach and note that when $r_H$ and $r_{p_T}$
are kept fixed we can separate massless cuts as in~\eqref{eq:Ftri_splitting}
and again end up with functions that are analytic in $z$ apart from a branch
cut for real $z>1$. Therefore it is possible to approximate the top-quark
mass dependence of the form factors at a given phase-space point, i.e. for
fixed $m_H^2$, $\hat{s}$ and $p_T^2$, by constructing a Pad\'e approximant that
describes the dependence on the variable $z$.

We find that the inclusion of the top threshold terms, as described for
the triangle form factor~\eqref{eq:Ftri_splitting} in Section~\ref{sec:ggH_Pade},
is of even greater importance for the construction of Pad\'e approximants for the
form factors~\eqref{eq:FiGGHH} than for $F_\triangle$. The computation of these 
terms is described in the following Section~\ref{sec:threshold} and our results 
are given in Appendix~\ref{sec:ggHH_Results}. Readers who are mostly interested 
in the phenomenological aspects may prefer to proceed to
Section~\ref{sec:numerics}. There, we assess the reliability of our
approach for Higgs pair production by performing a detailed comparison
with the exact NLO results.


\section{The amplitude near threshold\label{sec:threshold}}

In this section the computation of the non-analytic terms in the threshold
expansion of the form factors defined in Section~\ref{sec:method} is
described. Factorization formulae for the inclusive production cross section
of heavy-particle pairs near threshold have been developed in
\cite{Beneke:2003xh,Beneke:2004km,Beneke:2009rj,Beneke:2010da,Falgari:2012hx}
and applied to a number of processes~\cite{Beneke:2007zg,Actis:2008rb,Beneke:2010mp,
Jantzen:2013gpa,Beneke:2017rdn,Beneke:2011mq,Beneke:2012wb,Beneke:2016kvz}.
The approach is based on the factorization of forward-scattering amplitudes
which are related to the inclusive cross section by the optical theorem.
We have extended the factorization formula to the
$gg\to H^{(*)},HH,HZ,ZZ$ amplitudes. Only the basic aspects are sketched here
and the reader is referred to the original
literature~\cite{Beneke:2003xh,Beneke:2004km,Beneke:2009rj,Beneke:2010da,Falgari:2012hx}
for a detailed derivation and discussion.

\subsection{Structure of the amplitude near threshold\label{sec:structure}}

\begin{figure}
 \begin{center}
    \includegraphics[width=0.9\textwidth]{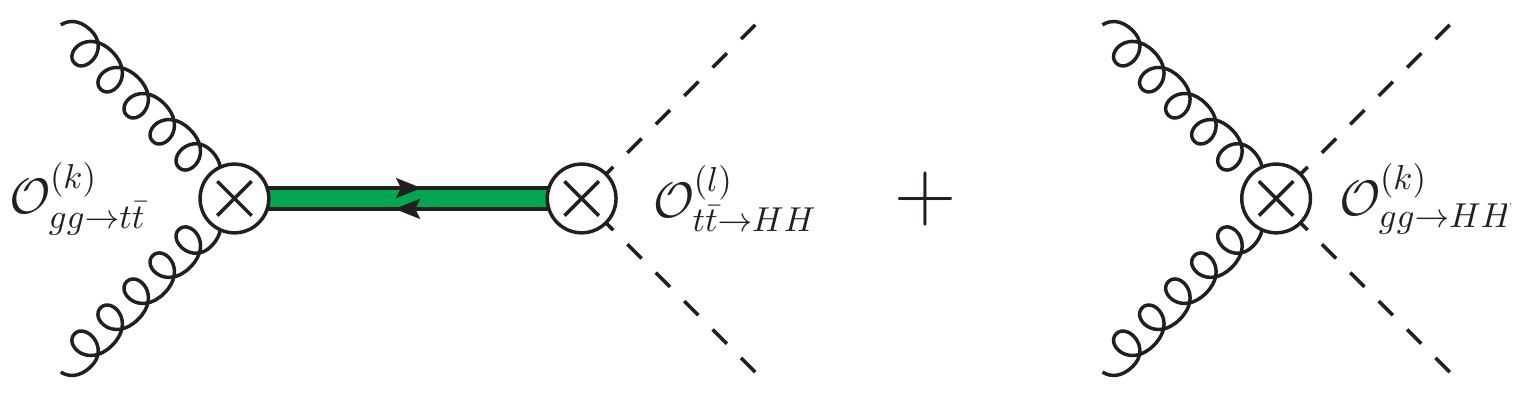}
  \caption{\label{fig:Factorization}
  Graphical representation of the terms in the master formula~\eqref{eq:Master_eq}.
  The diagram on the left (right) corresponds to the 'resonant' ('non-resonant')
  part of the amplitude. The shaded area indicates that Coulomb exchanges
  between the top quark pair are resummed.}
 \end{center}
\end{figure}

Near the threshold, $z\to1$, the top quarks can only be on shell if they
are non-relativistic. This implies a large hierarchy between the top mass
$m_t$, its typical momentum $m_t\sqrt{1-z}$ and its kinetic energy $m_t(1-z)$
which set the hard, soft and ultrasoft scale, respectively. Therefore, an
effective field theory (EFT) can be constructed by integrating out the hard
and soft scale. Then, the only dynamical modes left are non-relativistic
top quarks, collinear and ultrasoft gluons and the external fields.
The EFT describes the interactions of the remaining modes and is based on
\emph{potential non-relativistic QCD}
(PNRQCD)~\cite{Pineda:1997bj,Pineda:1997ie,Beneke:1998jj,Beneke:1999qg,Brambilla:1999xf,Beneke:2013jia}
and \emph{Soft Collinear Effective Theory} (SCET)~\cite{Bauer:2000ew,Bauer:2000yr,
Bauer:2001yt,Beneke:2002ph,Beneke:2002ni,Becher:2014oda}.
The amplitudes for $gg\to F$ with final states $F=H^{(*)},HH,HZ,ZZ$ are
given by the master
formula~(cf.~\cite{Beneke:2003xh,Beneke:2004km})
\begin{eqnarray}
 i\mathcal{A}_{gg\to F} & \mathop{=}\limits^{z\to1} & \sum\limits_{k,l}C_{gg\to t\bar{t}}^{(k)}\,C_{t\bar{t}\to F}^{(l)}\,
 \int d^4x\Braket{F|T\left[i\mathcal{O}_{t\bar{t}\to F}^{(l)}(x)i\mathcal{O}_{gg\to t\bar{t}}^{(k)}(0)\right]|gg}_\text{EFT} \nonumber \\
 & & + C_{gg\to F}\Braket{F|i\mathcal{O}_{gg\to F}(0)|gg}_\text{EFT},
 \label{eq:Master_eq}
\end{eqnarray}
where the matrix elements have to be evaluated in the EFT. In analogy
with~\cite{Beneke:2003xh,Beneke:2004km} we call the contributions in
the first and second line of~\eqref{eq:Master_eq} line the 'resonant'
and 'non-resonant' amplitude, respectively. This structure is shown
in Figure~\ref{fig:Factorization} in diagrammatic form.

The 'resonant' part in the first line of~\eqref{eq:Master_eq} contains
the contributions that involve a non-relativistic top quark pair, i.e. a top
pair that is close to being on resonance. This entails that only a soft spatial
momentum can be exchanged between the initial and final state. Since the
incoming gluons contain hard momentum components they must be
connected by a hard subgraph. The same holds for the two final state
particles. Integrating out these hard subgraphs yields local production
operators
\begin{equation}
 \left[\mathcal{O}_{gg\to t\bar{t}}^{(k)}\right]^{\mu\nu} = \mathcal{A}_c^{\perp\mu}\mathcal{A}_{\bar{c}}^{\perp\nu}\,\psi^\dagger\Gamma^{(k)}\chi,
 \label{eq:production_op}
\end{equation}
that annihilate the incoming gluons and create a non-relativistic top
pair and local annihilation operators
\begin{equation}
 \mathcal{O}_{t\bar{t}\to F}^{(l)}  = \chi^\dagger\Gamma^{(l)}\psi\,\phi_F^\dagger,
 \label{eq:annihilation_op}
\end{equation}
that annihilate the top pair and create the final-state particles.
Here $\mathcal{A}_{\bar{c}}^{\perp}$ is the collinear gluon field given
in~\cite{Beneke:2010da}, the non-relativistic two-component spinor
fields $\psi$ and $\chi$ annihilate a top quark and produce an anti-top
quark respectively, $\Gamma^{(k)}$ contains a combination of Pauli matrices,
$SU(3)_c$ generators and potentially covariant derivatives and $\phi_F^\dagger$
represents a combination of fields that produces the final state.
Both types of operators have associated hard-matching coefficients
that absorb the higher-order corrections from hard modes. The propagation
of the non-relativistic top pair is subject to a non-local color Coulomb
interaction that manifests as $\alpha_s/\sqrt{1-z}$ corrections in the
amplitude. These so-called Coulomb singularities can be resummed to all
orders within PNRQCD. The 'resonant' contribution contains non-analytic
$\sqrt{1-z}$ and $\ln(1-z)$ terms that correspond to on-shell cuts of
the non-relativistic top pair.

Contributions where a hard momentum component is exchanged between the initial
and the final state are contained in the 'non-resonant' part in the second
line of~\eqref{eq:Master_eq}.
In the EFT they are represented by the matrix element of the local operator
\begin{equation}
 \left[\mathcal{O}_{gg\to F}\right]^{\mu\nu} = \mathcal{A}_c^{\perp\mu}\mathcal{A}_{\bar{c}}^{\perp\nu}\,\phi_F^\dagger,
 \label{eq:nonres_op}
\end{equation}
that annihilates the incoming state and creates the final state.
Since the top quarks cannot be on shell near threshold when they carry
hard momentum, there are no discontinuities associated with $t\bar{t}$ cuts.
Therefore, this contribution admits the form of a Taylor expansion in
$(1-z)$ once massless cuts have been separated as described in
Section~\ref{sec:ggH_Pade}. The computation of this contribution is
very involved since already the leading term in the Taylor expansion
has the complexity of the full amplitude evaluated directly at the
threshold $z=1$. However, we expect the Pad\'e approximation to predict
this unknown analytic part of the amplitude very accurately, even when using
only the LME as input. Indeed, as we showed explicitly in
Section~\ref{sec:ggH_Pade}, adding the knowledge of just the
non-analytic terms near threshold is already sufficient to reconstruct
the full top-quark mass dependence with high accuracy. Therefore we can
safely ignore the non-resonant contribution and only focus on the much
simpler factorizable part.

\subsection{Computation of the non-analytic terms\label{sec:computation}}

\begin{figure}
 \begin{center}
    \includegraphics[width=0.8\textwidth]{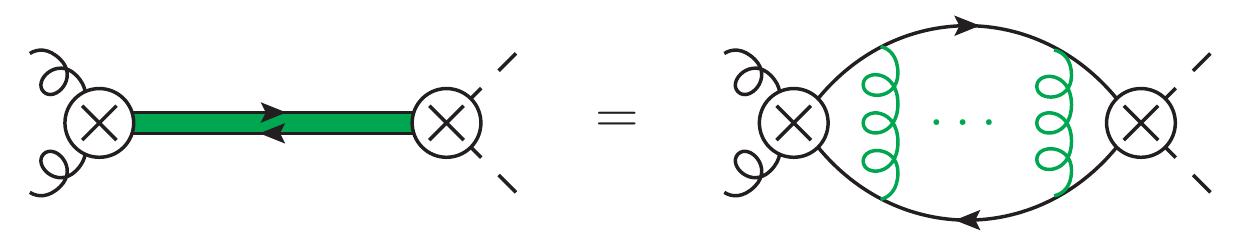}
  \caption{\label{fig:LO}
  Matrix element at leading order in the power counting $\alpha_s\sim\sqrt{1-z}$.}
 \end{center}
\end{figure}

In this section we describe the computation of the 'resonant' part of the 
amplitude~\eqref{eq:Master_eq}. We adopt here the non-relativistic power 
counting where $\alpha_s\sim\sqrt{1-z}$ and denote the $k$'th order in 
this counting by nrN$^k$LO to distinguish it from the fixed-order expansion 
in the strong coupling constant. At nrLO, the matrix element is given by 
a non-relativistic Green function which resums the $1/\sqrt{1-z}$ enhanced 
effects from the ladder-exchange of Coulomb gluons as indicated in 
Figure~\ref{fig:LO}. Hence, at any loop order, the leading non-analytic 
term in the threshold expansion of the amplitude can be determined by 
expanding the nrLO result to the respective order in $\alpha_s$. 
Up to nrNNLO, terms of the relative order
\begin{equation}
 \frac{\mathcal{A}_\text{'resonant'}}{\mathcal{A}_0(z=1)} \sim \sqrt{1-z}^{\,2l+1} \sum\limits_{k=0}^\infty \left(\frac{\alpha_s}{\sqrt{1-z}}\right)^k
 \times \begin{cases}
         \begin{array}{ll}
          1 & \text{nrLO},\\
          \alpha_s,\sqrt{1-z} & \text{nrNLO},\\
          \alpha_s^2,\alpha_s \sqrt{1-z}, (1-z)\hspace{1cm} & \text{nrNNLO},
         \end{array}
        \end{cases}
 \label{eq:scaling}
\end{equation}
must be included, where $\mathcal{A}_0(z=1)$ is the LO amplitude evaluated 
\emph{at} the top threshold, $l=0,1,\dots$ denotes the angular momentum
of the top pair and the global factor $\sqrt{1-z}$ accounts for
the suppression of the phase-space near threshold.

\begin{figure}[t]
 \begin{center}
    \includegraphics[width=0.78\textwidth]{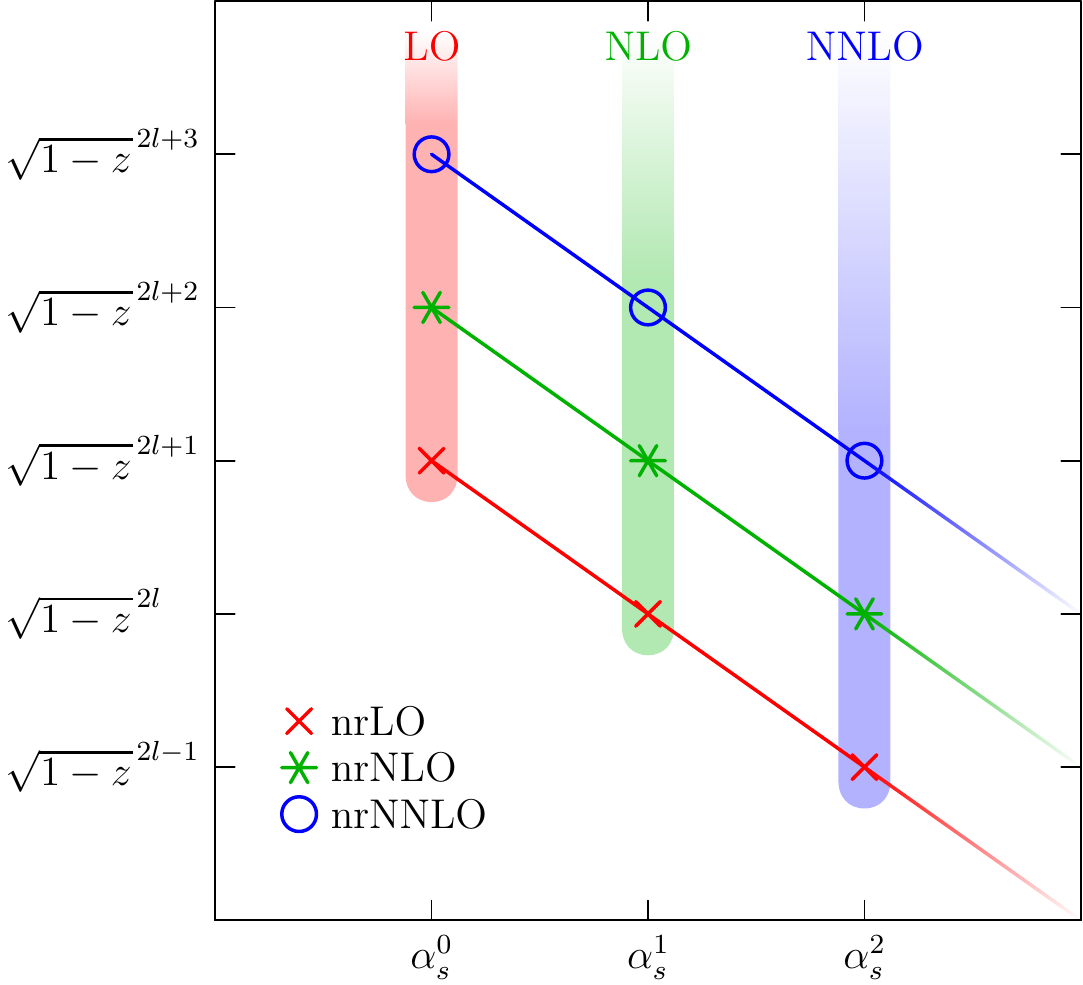}
  \caption{\label{fig:poco}
  Relation between relativistic (LO, NLO, NNLO) and non-relativistic
  (nrLO, nrNLO, nrNNLO) power counting up to next-to-next-to-leading
  order. The axes show the powers of $\alpha_s$ and $\sqrt{1-z}$ in the
  various coefficients represented by the markers. Note that the normalization
  is chosen such that $\alpha_s^0$ corresponds to LO.}
 \end{center}
\end{figure}

Fig.~\ref{fig:poco} illustrates the relation between different orders in
standard relativistic perturbation theory and in the non-relativistic
effective theory. For example, the
following terms on the right-hand side of Eq.~\eqref{eq:scaling}
contribute to the fixed-order expansion up to NLO:
\begin{itemize}
\item The nrLO terms with relative factors $\sqrt{1-z}^{\,2l+1},
  \alpha_s\sqrt{1-z}^{\,2l}$.
\item The nrNLO terms with relative factors $\sqrt{1-z}^{\,2l+2},
  \alpha_s\sqrt{1-z}^{\,2l+1}$.
\item The nrNNLO terms with relative factors $\sqrt{1-z}^{\,2l+3},
  \alpha_s\sqrt{1-z}^{\,2l+2}$.
\end{itemize}

For the processes $gg\to H^{(*)}$ and $gg\to HH$ there is no
contribution from S-wave $t\bar{t}$ states due to parity and C-parity
conservation.\footnote{The $H$ and $HH$ final states have even parity and
C-parity and the $t\bar{t}$ state with angular momentum $l$ and spin $s=0,1$
has $P=(-1)^{l+1}$ and $C=(-1)^{l+s}$. Thus, $l$ is one ($H$) or odd
($HH$) and $s=1$.} The leading 'resonant' contribution therefore contains
the P-wave Green function~\cite{Beneke:2013kia} which is suppressed by
$(1-z)$ near threshold.
We want to determine the 'resonant' amplitude up to nrNLO in the
scaling~\eqref{eq:scaling}, which contains the next-to-leading non-analytic 
terms in the threshold expansion at any loop order. In addition we compute 
the first two terms in the fixed-order expansion of the nrNNLO result in
$\alpha_s$, i.e. those of relative orders $(1-z)^{5/2}$ and $\alpha_s(1-z)^2$.
They correspond to the next-to-next-to leading threshold terms for the one 
and two loop amplitude which we study in Section~\ref{sec:numerics}.

\begin{figure}
 \begin{center}
    \includegraphics[width=0.45\textwidth]{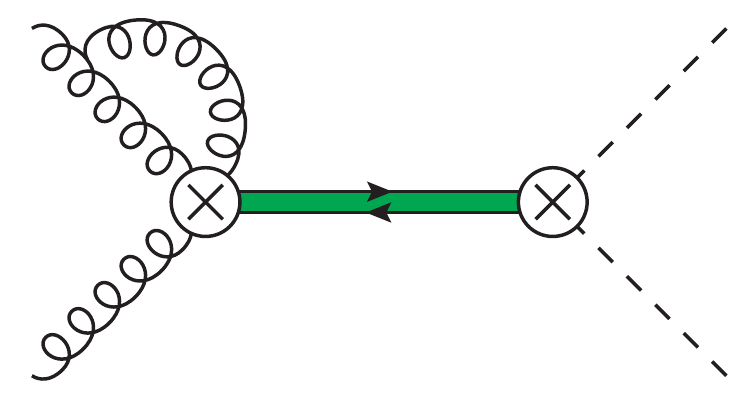}
    \includegraphics[width=0.45\textwidth]{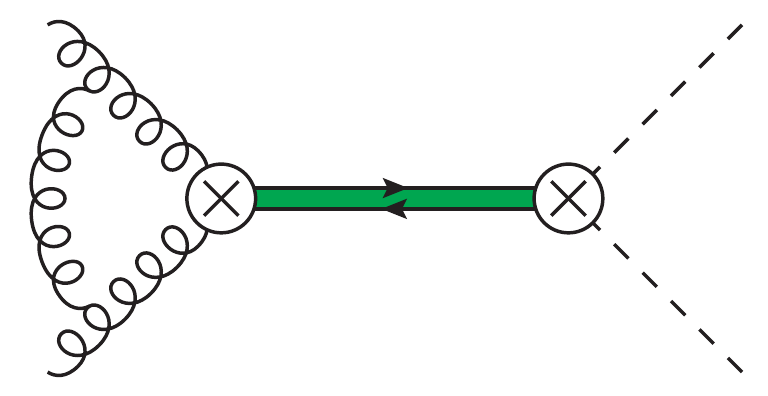}
  \caption{\label{fig:CollinearUltrasoft}
  nrNLO diagrams involving collinear (left) and ultrasoft (right) gluon radiation.
  Both loop integrals are scaleless and vanish in dimensional regularization.}
 \end{center}
\end{figure}

The matrix elements in~\eqref{eq:Master_eq} receive corrections from 
the higher-order
non-local potentials and the dynamical modes contained in the EFT.
The EFT contains no interactions of collinear modes with
non-relativistic modes or between collinear modes of different
directions. They cannot be present because the combination of the
involved momenta yields hard modes which have been integrated out.
Therefore the only collinear corrections at nrNLO are from the left
diagram in Figure~\ref{fig:CollinearUltrasoft}. The corresponding
loop integral is scaleless and therefore vanishes in dimensional
regularization.

Ultrasoft gluons couple to the collinear and non-relativistic sector as
well as to the P-wave production and annihilation operators.
The exchange of ultrasoft gluons between the collinear states shown in
the diagram on the right of Figure~\ref{fig:CollinearUltrasoft} yields
only scaleless integrals.
The interactions in the EFT must be multipole expanded. At leading order 
in the multipole expansion
ultrasoft gluons couple to the net color charge of the $t\bar{t}$ state
since the large wavelength $\lambda\sim1/(m_t(1-z))$ gluons cannot
resolve the spatial separation $a_\text{B}\sim1/(m_t\sqrt{1-z})$ of the top pair.
The first non-vanishing term in the multipole expansion for color
singlet states is therefore the chromoelectric term
$\psi^\dagger\,\mathbf{x}\cdot\mathbf{E}\,\psi$ which is suppressed
by $\alpha_s^{1/2}\sqrt{1-z} \sim \alpha_s^{3/2}$.
Similarly the ultrasoft gluon term in the covariant derivative in the
P-wave operators is suppressed by $\alpha_s^{1/2}\sqrt{1-z} \sim \alpha_s^{3/2}$
with respect to the derivative term. A single insertion of either of
these subleading terms vanishes by rotational invariance~\cite{Falgari:2012hx}.
Thus, contributions from the subleading interactions require at least
two insertions and first appear at nrNNNLO.

The effects of higher-order potentials enter as corrections to the
non-relativistic Green function. The nrNNNLO S-wave and nrNLO P-wave Green
functions have been computed for $t\bar{t}$ production in $e^+e^-$
collisions near threshold \cite{Beneke:2015kwa,Beneke:2013kia}. We
determine the $\alpha_s^{0,1}$ terms in the nrNNLO P-wave Green function
in Appendix~\ref{sec:Pwave}. Up to the considered order the resonant
amplitudes hence take the simple factorized form
\begin{equation}
 \mathcal{A}_\text{resonant} = \sum\limits_{k,l}\,\mathcal{N}_{kl}(1-z)\,C_{gg\to t\bar{t}}^{(k)}\,C_{t\bar{t}\to F}^{(l)}\,G_{S,P}(1-z).
 \label{eq:resonant_factorization}
\end{equation}
The Wilson coefficients $C_{gg\to t\bar{t}}^{(k)},\,C_{t\bar{t}\to F}^{(l)}$ 
are perturbative in $\alpha_s$ and independent of~$z$. We can compute them 
via matching to the full Standard Model, i.e.~by performing a Taylor 
expansion of the on-shell amplitudes for $gg\to t\bar{t},\, t\bar{t}\to F$ 
around the top threshold and comparing to the matrix elements of the effective 
operators $\mathcal{O}_{gg\to t\bar{t}}^{(k)},\,\mathcal{O}_{t\bar{t}\to F}^{(l)}$. 
Subleading terms in the Taylor expansion in $(1-z)$ correspond to 
higher-dimensional operators, which contain derivatives acting on the 
non-relativistic top and anti-top fields. Since $(1-z)\sim\alpha_s^2$, 
we only require matrix elements with at most one subleading operator 
up to nrNNLO. The normalization factors $\mathcal{N}_{kl}$ are either 
$z$-independent, if the operators $\mathcal{O}_{gg\to t\bar{t}}^{(k)}$
and $\mathcal{O}_{t\bar{t}\to F}^{(l)}$ are of leading order in the 
non-relativistic expansion, or proportional to $(1-z)\sim\alpha_s^2$, 
if one of the operators is of subleading order.  To achieve the
accuracies specified in~\eqref{eq:scaling} we require the following ingredients  
\begin{itemize}
 \item nrLO: 
   \begin{itemize}
   \item the tree-level coefficients $C_{gg\to t\bar{t}}^{(k)},\,C_{t\bar{t}\to F}^{(l)}$
   \item the nrLO Green function $G_{S,P}(1-z)$
   \end{itemize}
 \item nrNLO: the above and 
   \begin{itemize}
   \item the one-loop coefficients $C_{gg\to t\bar{t}}^{(k)},\,C_{t\bar{t}\to F}^{(l)}$
   \item the nrNLO Green function $G_{S,P}(1-z)$
   \end{itemize}
 \item the order $\alpha_s^{0,1}$ terms at nrNNLO: the above and 
   \begin{itemize}
   \item the tree-level coefficients $C_{gg\to t\bar{t}}^{(k)},\,C_{t\bar{t}\to F}^{(l)}$ 
   for the $(1-z)$-suppressed operators 
   \item the $\alpha_s^{0,1}$ terms in the nrNNLO Green function $G_{S,P}(1-z)$
   \end{itemize}
 \item nrNNLO: the above and 
   \begin{itemize}
   \item the two-loop coefficients $C_{gg\to t\bar{t}}^{(k)},\,C_{t\bar{t}\to F}^{(l)}$
   \item the nrNNLO Green function $G_{S,P}(1-z)$
   \end{itemize}
 \end{itemize}
As mentioned before, it is sufficient to know the nrNNLO terms
proportional to $\alpha_s^0$ and $\alpha_s^1$ in order to construct
approximations to two-loop (NLO) fixed-order amplitudes
(cf. Fig~\ref{fig:poco}). The remaining nrNNLO terms of the relative
order $\alpha_s^2(1-z)^{3/2}$ will be important for the reconstruction
of the three-loop amplitude. Since its determination requires the
calculation of the two-loop matching coefficients $C_{gg\to
t\bar{t}}^{(k)}$ and $C_{t\bar{t}\to F}^{(l)}$ as the most complicated
ingredient, we postpone this to future work.

The one-loop coefficients $C_{t\bar{t}\to F}^{(l)}$ are finite after field and
mass renormalization. The one-loop coefficients $C_{gg\to t\bar{t}}^{(k)}$,
however, require additional IR subtractions since the virtual amplitude
by itself is not IR safe. Our results for the threshold expansion of the
form factors are given in~\eqref{eq:triangle_threshold1}--\eqref{eq:triangle_threshold4} and
Appendix~\ref{sec:ggHH_Results} together with the details of the
IR subtractions. Together with the nrNLO expression for the P-wave
Green function~\cite{Beneke:2013kia} these results are sufficient to
determine the leading and next-to-leading non-analytic terms in the 
threshold expansion of the form factors at \emph{any order} in
$\alpha_s$.

Another interesting, yet more involved, application of our formalism
is Higgs plus jet production. Here, we shortly comment on that,
but leave a more careful assessment to future work.
The amplitudes $gg\to Hg$, $gq\to Hq$ and $q\bar{q}\to Hg$
obey the same structure of~\eqref{eq:Master_eq} near the top threshold
but the corresponding 'resonant' matrix elements are more
complicated since the final state now contains a color-charged particle.
Ultrasoft gluons can then be exchanged between the initial
state,  the final state and the intermediate top pair which is in a
color octet state and no longer decouples.
In \cite{Beneke:2009rj,Beneke:2010da,Falgari:2012hx} it was demonstrated
for arbitrary color structures that the 'resonant' matrix elements in
forward-scattering amplitudes factorize into the convolution of a
non-relativistic Green function,
therein called the potential function, and an ultrasoft function, therein
called the soft function. At leading power this follows from field transformations
that decouple the collinear and non-relativistic fields from the
ultrasoft fields. The extension to higher orders requires a careful
assessment of the subleading interactions and was performed to NNLL
in~\cite{Beneke:2009rj,Beneke:2010da,Falgari:2012hx}.
Following these derivations we identified no aspect that would obstruct
the extension to Higgs plus jet production and therefore conjecture
that an analogous factorization formula holds for the corresponding
amplitudes.

\section{Comparison with the exact result\label{sec:numerics}}
As a proof of method, we compare our results at LO and NLO with the results in full
top mass dependence for Higgs pair production.
While at LO, the Higgs pair production cross section is known in full mass dependence
since the late 80's \cite{Eboli:1987dy,Glover:1987nx, Plehn:1996wb}, the computation
of the NLO QCD corrections is quite involved, due to the many scales of the problem.
The first work on the NLO corrections
was based on the heavy top mass limit \cite{Dawson:1998py} reweighted with the
matrix elements squared of the full LO results (HEFT).
The real corrections in full top mass dependence have been
 computed in \cite{Maltoni:2014eza, Frederix:2014hta}, while the virtual corrections have been kept in HEFT.
 The computation of the virtual corrections in full top mass dependence
 became available only recently in \cite{Borowka:2016ehy, Borowka:2016ypz}.
 \subsection{Numerical setup\label{sec:numerical_setup}}
For the numerical evaluation we choose a centre-of-mass energy of $\sqrt{s}=14\text{ TeV}$.
The Higgs boson mass has been set equal to $m_{H}=125\text{ GeV}$ and the top quark mass to $m_t=173\text{ GeV}$.
We do not account for bottom quark loops as they contribute with less than 1\% at LO.
 We have adopted the PDF set {\tt NNPDF3.0} \cite{Ball:2014uwa}. The strong coupling
 constant is set to $\alpha_s(M_Z)=0.118$ at LO and NLO. The renormalization scale
 has been set to $M_{HH}/2$, where $M_{HH}$ denotes the invariant mass of the Higgs
 boson pair, as suggested by the NNLL soft gluon resummation performed in \cite{Shao:2013bz,deFlorian:2015moa}.
\par
 We construct our Pad\'e approximants at LO (NLO) as described in
 Section~\ref{sec:method} by solving numerically the 8 (7) equations
 from the LME  \cite{Degrassi:2016vss} and threshold expansion, given in
 Section \ref{sec:ggH_Pade} and Appendix~\ref{sec:ggHH_Results}, 
 by means of the {\tt FORTRAN} routine {\tt MINPACK}~\cite{minpack}.\footnote{We
   provide a {\tt FORTRAN} routine of the Pad\'e approximated matrix
   elements upon request.}
For every phase space point we construct a total of 100 Pad\'e approximants
$[n/m]$, where  $a_{R}$ takes a random value between [0.1,10], $n,m \in [1,6]$
at LO and $n,m \in [1,5]$ at NLO, and take the mean value. From that we
obtain an error estimate on every form factor by taking the standard deviation.
For the computation of the cross section or the virtual corrections we add up
the errors stemming from the different form factors quadratically. Pad\'e
approximants with poles in $\text{Re}(z) \in [0,8]$ and $\text{Im}(z)\in
[-1,1]$ were excluded, since
functions with poles close-by in the complex plane could have an unwanted
resonant behaviour.
The running time per phase space point for the construction of
100 Pad\'e approximants at NLO is usually below 6 s.

 \subsection{Comparison at LO}
 In Table \ref{tab:LOcxn} we give the results for the LO cross section
 in different approximations. The first row, $[n/m]$ w/o THR, symbolizes
 the cross section obtained with Pad\'e approximants constructed without input
 from the threshold expansion, where $n,m \in [1,3]$ and approximants with
 poles as described above have been excluded. The result we obtain when
 including the threshold information and using the specifications described
 in Section~\ref{sec:numerical_setup} is denoted by $[n/m]$. With $[n/n\pm 1, 3]$
 we symbolize the results we find when only the Pad\'e approximants
 [5/2], [4/3], [3/4] and [2/5] are used.\footnote{
 Note however that these are mainly [5/2] and [4/3] Pad\'e approximants as
 [3/4] and [2/5] usually are excluded by our pole criterion.}
 Finally, we give the full LO cross section (obtained with {\tt HPAIR}
 \cite{hpair}) in the fourth row of Table \ref{tab:LOcxn}.
As can be inferred from the table, the Pad\'e approximants provide a
very good approximation for the full cross section, in particular if only
the most diagonal and next-to-diagonal Pad\'e approximants are constructed.
The threshold expansion proves to be essential for a good approximation.
As expected, the standard deviation computed from the construction of 100
$[n/m]$ Pad\'e approximants with random $a_R$ and different $n,m$
becomes smaller if we construct only the most diagonal and next-to-diagonal
Pad\'e approximants.
 \begin{table}
 \begin{center}
\renewcommand{\arraystretch}{1.2}
 \begin{tabular}{cc}
 \toprule
  & $\sigma$ [fb] \\
\midrule
 $[n/m]$ w/o THR & $19.9\pm 5.4$ \\
 $[n/m]$ & $21.7 \pm 1.1$\\
 $[n/n\pm 1,3] $& $ 21.3 \pm 0.4 $ \\
 full & 21.3 \\ \bottomrule
 \end{tabular}
 \end{center}
 \caption{Numbers for the total LO cross section and standard deviation
   from the construction of 100 Pad\'e approximants. \label{tab:LOcxn}
   }
 \end{table}
  \begin{figure}[ht!]
 \centering
\includegraphics[width=14cm]{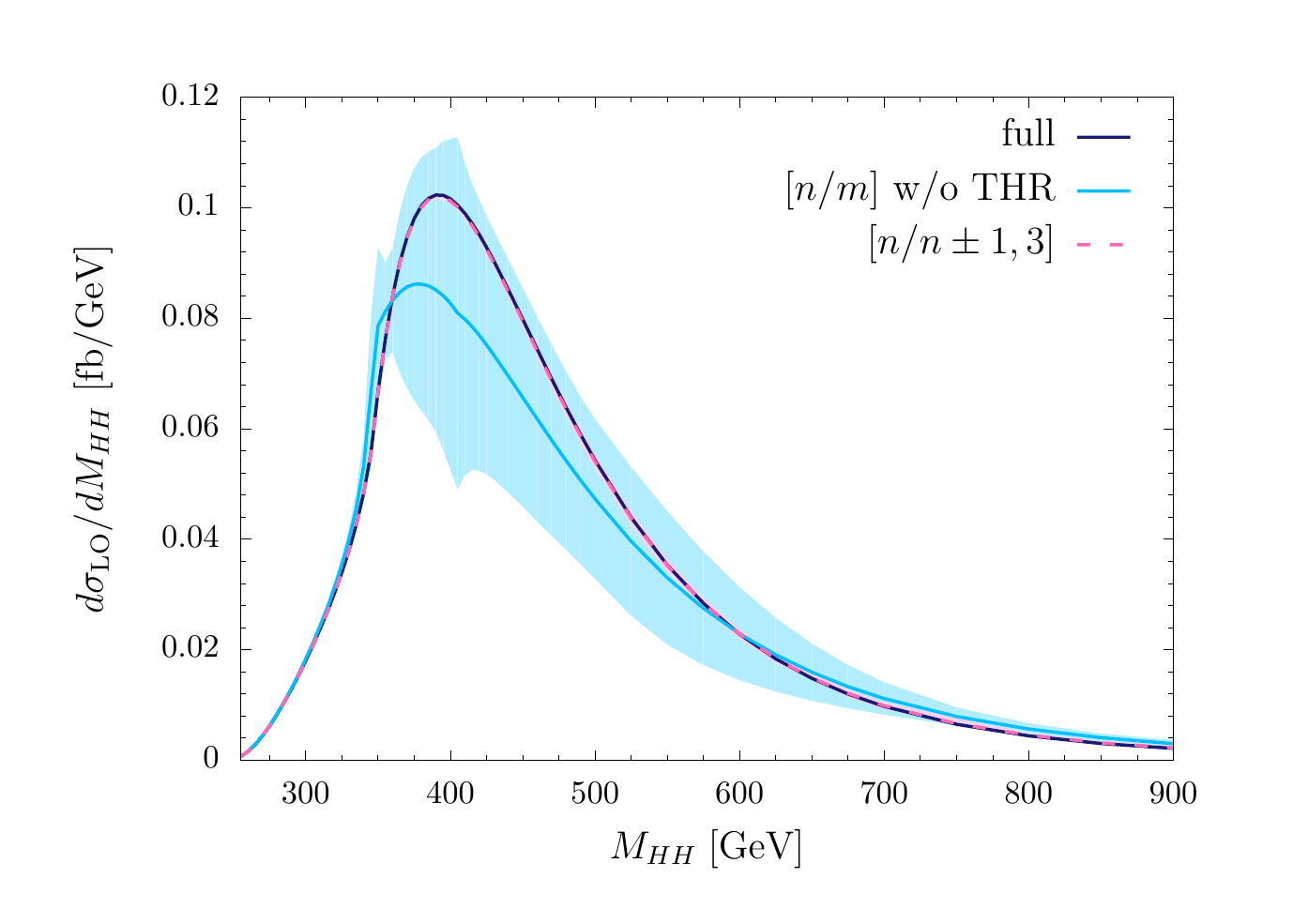}
\caption{Invariant Higgs mass distribution for the full LO cross section
  (dark blue), the $[n/ n\pm 1,3]$ Pad\'e approximants (pink line) and
  the Pad\'e constructed without threshold expansion (light blue). The
  standard deviation of the Pad\'e lines are shown by the
  semi-transparent regions with the corresponding color. The pink 
  band is barely wider than the width of the curves and hardly 
  visible.\label{fig:LOMHH}}
 \end{figure}
 \par
 In Fig.~\ref{fig:LOMHH} we show the invariant Higgs mass distribution for
 the full result (dark blue), the $[n/n \pm 1,3] $ Pad\'e approximants (pink)
 and the Pad\'e approximants without the threshold expansion (light blue).
 While the $[n/n\pm 1,3] $ full Pad\'e approximants fit the shape of the
 invariant mass distribution in full mass dependence almost perfectly,
 the approximation where the threshold expansion is not included
 (hence the approximation is only built from the LME) fits the shape only
 for small invariant mass. The error on the construction of the approximation
 including the threshold expansion is rather small whereas if the approximation
 is constructed only from the LME, the error becomes much larger in particular
 above the threshold.
 \par
 We thus conclude that at LO our approximation of the mass effects by
 Pad\'e approximants works well as long as the conditions obtained from
 the threshold expansion are included. Using only nearly diagonal
 Pad\'e approximants leads to a result with smaller error
 with values closer to the true result.

\subsection{Comparison at NLO}
 Finally, we compare our results to the computation of the NLO corrections in
 full top mass dependence of Ref.~\cite{Borowka:2016ehy, Borowka:2016ypz}.
 In the framework of Ref.~\cite{Heinrich:2017kxx} a grid and an interpolation
 function with numerical values for the virtual corrections of
 Ref.~\cite{Borowka:2016ehy, Borowka:2016ypz} have been provided.
 \\
 In order to fit the conventions of Ref.~\cite{Heinrich:2017kxx} we define the finite part of the virtual corrections as
 \begin{equation}
 \begin{split}
 \mathcal{V}_{fin}=&\frac{\alpha_s^2(\mu_R)}{16 \pi^2}\frac{\hat{s}^2}{128 v^4}\Bigg[\left|\mathcal{M}_{born}\right|^2\left(C_A \pi^2 - C_A \log^2\left(\frac{\mu_R^2}{\hat{s}}\right)\right)\\ 
 +& 2 \left\{(F_1^{1l})^*\left(F_1^{2l,[n/m]}+F_1^{2\Delta}\right)+(F_2^{1l})^*\left(F_2^{2l, [n/m]}+F_2^{2\Delta}\right)+\text{h.c.}\right\}\Bigg]
 \end{split}
 \end{equation}
with
\begin{equation}
\left|\mathcal{M}_{born}\right|^2=\left|F_1^{1l} \right|^2  + \left|F_{2}^{1l}\right|^2
\end{equation}
and $F_1$ defined in eq.~\eqref{eq:f1}. For $F_x^{2l,[n/m]}$ we use the
matrix elements constructed with the Pad\'e approximant $[n/m]_{\tilde{f}}$.
All other matrix elements are used in full top mass dependence. The form
factors $F_x^{2\Delta}$ stem from the double triangle contribution to the
virtual corrections and can be expressed in terms of one-loop integrals.
They are given in Ref.~\cite{Degrassi:2016vss} in full top mass dependence.
In the heavy top mass limit they become
\begin{equation}
F^{2\Delta}_1\to \frac{4}{9}, \hspace*{0.5cm} F^{2\Delta}_2\to -\frac{4}{9}\frac{p_T^2}{2\hat{t}\hat{u}}(\hat{s}- 2 m_H^2).
\end{equation}
 The contribution of the double triangle diagrams to the virtual
 corrections is only of the order of a few per cent \cite{Grober:2015cwa}.
 \begin{table}
\begin{center}
\renewcommand{\arraystretch}{1.15}
\begin{tabular}{cr@{.}lc  c c c c}
  \toprule
  \multicolumn{3}{c}{} &     \multicolumn{4}{c}{$\mathcal{V}_{fin}\times 10^4$}\\ \cmidrule(l){4-7}
  $M_{HH} $\,[GeV] & \multicolumn{2}{c}{$p_T$\,[GeV]}&
                                                 HEFT & $[n/m]$ & $[n/n\pm 0, 2]$ & full \\
  \midrule
336.85 &  37 & 75 & $0.912$ & $0.996 \pm 0.004$ & $0.990 \pm 0.001$ & $0.996 \pm 0.000$ \\
350.04 & 118 & 65 & $1.589$ & $1.933 \pm 0.012$ & $1.937 \pm 0.010$ & $1.939 \pm 0.061$ \\
411.36 & 163 & 21 & $4.894$ & $4.326 \pm 0.183$ & $4.527 \pm 0.069$ & $4.510 \pm 0.124$ \\
454.69 & 126 & 69 & $6.240$ & $5.300 \pm 0.192$ & $5.114 \pm 0.051$ & $5.086 \pm 0.060$ \\
586.96 & 219 & 87 & $7.797$ & $4.935 \pm 0.583$ & $5.361 \pm 0.281$ & $4.943 \pm 0.057$ \\
663.51 &  94 & 55 & $8.551$ & $5.104 \pm 1.010$ & $4.096 \pm 0.401$ & $4.120 \pm 0.018$ \\ \bottomrule
 \end{tabular}
\end{center}
 \caption{Numbers for the virtual corrections for some representative
   phase space points for the HEFT result reweighted with the full Born
   cross section (as in Ref.~\cite{Dawson:1998py}), the Pad\'e-approximated
   ones and the full calculation \cite{Heinrich:2017kxx}.  \label{tab:NLOVfin} }
 \end{table}
 \par
In Table~\ref{tab:NLOVfin} we compare values for the full computation of
the virtual corrections obtained from the grid of
Ref.~\cite{Heinrich:2017kxx}, the HEFT results rescaled with the full
Born cross section (as e.g.~implemented in {\tt HPAIR}), and the Pad\'e
approximations including all possible approximants without poles in
$\text{Re}(z) \in [0,8]$ and $\text{Im}(z)\in[-1,1]$ (called $[n/m]$)
and the ones where we only construct diagonal [3/3] and next-to diagonal
[4/2] and [2/4] approximants (called $[n/n\pm 0, 2]$). The errors
given in the table are, in case of the Pad\'e-approximated results, due
to the construction of the different approximants and due to the rescaling
with $a_R$. For the full results the error stems from internal binning
in the grid. As can be inferred from the table, the Pad\'e construction
approximates the full result quite well. It provides a much better
approximation than the HEFT results with a generally reliable error
estimate. While up to $M_{HH}=450$\text{ GeV} the Pad\'e method provides
an excellent approximation on the level of $\lesssim 2\%$, for larger
invariant masses and $p_T$ the results worsen gradually.
As already anticipated from the LO results, constructing only diagonal
and next-to diagonal Pad\'e approximants improves both the error and the
values of the virtual corrections with respect to the full
result. Indeed we even find that only constructing diagonal Pad\'e
approximants gives results even closer to the full result. Since this
does not allow for a reliable error estimate any more (the error would
then solely stem from the variation of $a_R$) we do not discuss this
here any further.
\par
 \begin{figure}[ht!]
 \centering
\includegraphics[width=14cm]{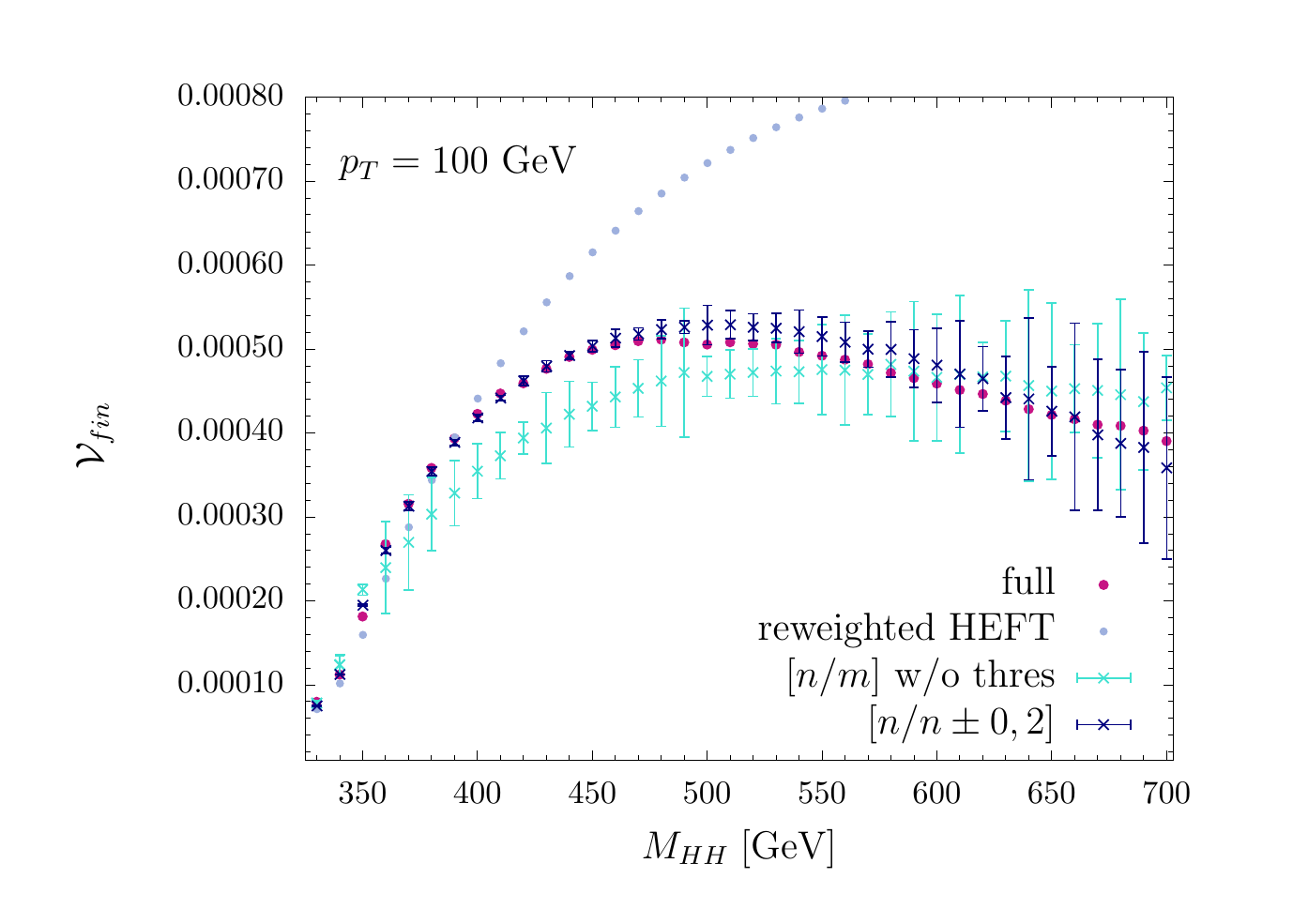}
\caption{Finite part of the virtual corrections, $\mathcal{V}_{fin}$, as
  a function of $M_{HH}$ for $p_T=100\text{ GeV}$. The light blue points
  are the reweighted HEFT results, the pink points the virtual
  corrections in full top mass dependence from the interpolation
  function provided with Ref.~\cite{Heinrich:2017kxx}, the dark blue
  points are from the diagonal and off-diagonal Pad\'e approximants with
  their standard deviation and the turquoise points with standard deviation are the Pad\'e approximants constructed
  without the threshold expansion. \label{fig:NLOcomp}}
 \end{figure}
In Fig.~\ref{fig:NLOcomp} we show for $p_T=100\text{ GeV}$ the virtual
corrections ${\cal V}_{fin}$ for varying $M_{HH}$ for the Pad\'e
approximations $[n/n\pm 0,2]$, the Pad\'e approximants 
constructed only from the LME, the full result and the reweighted HEFT
results.  
Again, we can see that contrary to the HEFT results the Pad\'e
approximation can reproduce the correct scaling with the invariant mass of the full result.  
The quality of the approximation is improved significantly with the
inclusion of the threshold expansion.
The error of the Pad\'e approximation increases with the invariant
mass. Note that the full result has, apart from the previous error from
the internal binning, also an error due to the interpolation
procedure. We do not quantify this error but in comparison to the HEFT
grid provided with Ref.~\cite{Heinrich:2017kxx} we conclude that while
in the range up to $M_{HH}\lesssim 570 \text{ GeV}$ this error is
negligible, it will be a few \% for larger $M_{HH}$. The comparison 
with the numerical results of~\cite{Heinrich:2017kxx} demonstrates that 
our prescription for the uncertainty related to the construction of Pad\'e 
approximants also provides a reasonable error estimate at NLO. 
\par
In conclusion, we see that for the NLO corrections the Pad\'e approximation
reproduces the correct scaling behaviour for small and moderate invariant
mass and $p_T$. Since the cross section peaks around $M_{HH}\approx 400\text{ GeV}$
and $p_T\approx 150 \text{ GeV}$ this will lead to a reliable approximation
and reliable error estimate also for the full cross section. It can be expected
that both the error and the difference with respect to the full result improves
once more input is used (i.e. higher orders in the threshold expansion,
higher orders in the LME, possibly input from a small mass expansion).

\section{Conclusions and outlook\label{sec:conclusion}}

We have reconstructed the top-quark mass dependence of the one and two loop
virtual amplitudes for Higgs pair production in gluon fusion with Pad\'e
approximants based on the LME of the amplitude~\cite{Degrassi:2016vss} and
new analytic results near the top threshold $\hat{s}=4m_t^2$. We observe perfect
agreement of the one-loop results with the exact expressions once the additional
conditions from the threshold terms are imposed. Significant deviations are
observed when only the LME is used to construct Pad\'e approximants, but
we still find agreement within the uncertainty estimate of our reconstruction,
which is based on variation of the rescaling parameter $a_R$ and the use of
different $[n/m]$ approximants. At the two-loop level the full result can be reproduced
in the entire phenomenologically relevant range within typical uncertainties
ranging from below $\pm 3\%$  in the region $M_{HH}\leq450$ GeV up to about $\pm 20\%$
for $M_{HH}=700$ GeV. Thus, our method allows for a determination of the
total cross section including top-quark mass effects at NLO where the uncertainty
due to the reconstruction is negligible compared to the scale uncertainty
which is of the size of $\pm13$\%~\cite{Borowka:2016ehy, Borowka:2016ypz}.
This represents considerable progress compared to the rescaled HEFT and LME
approximations where a reliable uncertainty estimate is not possible.
Our method can also be systematically improved by including higher orders
in the LME or threshold expansions. We expect even better behaviour if
one also considers the leading term in the small-mass expansion $z\to\infty$
which corresponds to the bottom-quark contribution expanded for small $m_b$.
An approach for computations in this limit has recently been introduced
\cite{Mueller:2015lrx,Melnikov:2016qoc,Melnikov:2017pgf}.
Furthermore our results strongly suggest that the combination of the Pad\'e
approximants of the NLO virtual corrections with the exact evaluation of the
real corrections~\cite{Maltoni:2014eza, Frederix:2014hta} can reproduce
differential distributions to high accuracy.

There is a large number of possible applications for our method.
To further increase the precision for Higgs pair production one needs to
consider NNLO QCD corrections. The rescaled HEFT approximation for the NNLO
corrections increases the cross section by 18\%~\cite{Borowka:2016ypz} which
exceeds the estimate from scale variation at NLO. A NNLO computation which
retains the full top-quark mass effects is clearly out of reach of the
current technology. On the other hand, the LME has already been computed up
to $1/m_t^4$ in~\cite{Grigo:2015dia} and we have determined the two first 
non-analytic terms in the threshold expansion. This presently available input
only allows for the construction of Pad\'e approximants with $n+m=3$ where we do
not expect stable behaviour, but a calculation of two or three more
expansion parameters would allow the evaluation of NNLO corrections in the
soft-virtual approximation of~\cite{Grigo:2015dia,deFlorian:2012za}.
Additionally, one can study the NLO electroweak corrections involving
top-quark loops. Of particular interest are the contributions involving
additional Higgs bosons which alter the dependence of the cross section
on the values of the Higgs self couplings.

It is straightforward to apply our method to $gg \to HZ$ and the top-quark mediated $gg \to ZZ$ 
amplitude and at higher orders in perturbation theory. In all these cases, results in
the LME have been obtained at two loops~\cite{Hasselhuhn:2016rqt,Campbell:2016ivq,Caola:2016trd}
and for $gg\to H^{(*)}$ even at three
loops~\cite{Harlander:2009bw,Pak:2009bx,Harlander:2009mq,Pak:2009dg,Harlander:2009my}.
The determination of the threshold terms only requires the computation of the
respective one-loop matching coefficients in~\eqref{eq:resonant_factorization}.
Another phenomenologically very interesting case is Higgs plus jet production.
The construction of Pad\'e approximants is also possible here but the computation
of the threshold expansion is more involved as we outlined in Section~\ref{sec:computation}.
Beyond LME results, also the leading term in the small-mass expansion is know
for the relevant two-loop amplitudes~\cite{Melnikov:2016qoc,Melnikov:2017pgf}.
Hence, the effects of this additional input on the reconstruction of top-quark
mass effects can be studied in this case.

\subsubsection*{Acknowledgements}
We thank Johannes Schlenk for useful discussions and Matthias Kerner for clarifications regarding the grid of Ref.~\cite{Heinrich:2017kxx}. RG acknowledges useful discussion with Fabrizio Caola, Keith Ellis and Sebastian Kirchner in an early stage of this project.
RG and AM are supported by a European Union COFUND/Durham Junior Research Fellowship under the EU grant number 609412.

\appendix

\section{Subtractions\label{sec:subtractions}}

We construct functions for the threshold subtractions based on the known
analytical results for the current correlators. The subtraction functions
and their threshold expansions are
\begin{eqnarray}
 s_1(z) & = & \frac{2}{\pi }\,(1-z) G(z)                     \,\,\mathop{\asymp}\limits^{z\to1}\,\, \sqrt{1-z} + \frac12 (1-z)^{3/2} + \frac38 (1-z)^{5/2} + \mathcal{O}\left((1-z)^{7/2}\right), \nonumber\\
 s_2(z) & = & -\frac{16 (1-z) \Pi^{(1),v}(z)}{3 z}           \nonumber \\
 & \mathop{\asymp}\limits^{z\to1} & (1-z)\ln(1-z) -\frac{8}{\pi} (1-z)^{3/2} + \frac13 (1-z)^2\ln(1-z) + \mathcal{O}\left((1-z)^{5/2}\right), \nonumber\\
 s_3(z) & = & \frac{2}{\pi}\,\frac{(1-z)^2 G(z)-1}{z}        \,\,\mathop{\asymp}\limits^{z\to1}\,\, (1-z)^{3/2} + \frac32 (1-z)^{5/2} + \mathcal{O}\left((1-z)^{7/2}\right), \nonumber\\
 s_4(z) & = & -\frac{8 }{81 \pi ^2}\,\frac{54 \pi ^2 (1-z)^2 \Pi^{(1),v}(z)-41 z}{z^2}      \,\,\mathop{\asymp}\limits^{z\to1}\,\, (1-z)^2\ln(1-z) + \mathcal{O}\left((1-z)^{5/2}\right), \nonumber\\
 s_5(z) & = & \frac{2}{3 \pi}\,\frac{3 (1-z)^3 G(z)+7 z-3}{z^2}                             \,\,\mathop{\asymp}\limits^{z\to1}\,\, (1-z)^{5/2} + \mathcal{O}\left((1-z)^3\right),
 \label{eq:subs}
\end{eqnarray}
where we have used the symbol $\asymp$ to denote that terms analytical
in $(1-z)$ have been dropped on the right-hand side,
\begin{equation}
 G(z) = \frac{1}{2z\sqrt{1-1/z}}\ln\left(\frac{\sqrt{1-1/z}-1}{\sqrt{1-1/z}+1}\right),
\end{equation}
and $\Pi^{(1),v}$ is the well-known two-loop correction to the vacuum
polarization~\cite{Kallen:1955fb} in the convention
of~\cite{Kiyo:2009gb}. The functions $s_i$ in~\eqref{eq:subs} are
constant as $z\to0$ and only diverge logarithmically as $z\to\infty$.

\section{Expansion of the P-wave Green function\label{sec:Pwave}}

The P-wave Green function has been computed up to nrNLO in~\cite{Beneke:2013kia}.
In addition we have determined the terms of order $\alpha_s^0$ and
$\alpha_s^1$ in the nrNNLO correction. Those are given by the insertion of the
'kinetic potential'~\cite{Beneke:2013jia}
\begin{equation}
 V_\text{kin}(\mathbf{p},\mathbf{p}') = -\frac{\mathbf{p}^4}{4m_t^3}(2\pi)^{d-1}\delta^{(d-1)}(\mathbf{p}-\mathbf{p}')
\end{equation}
and the $1/m^2$ potential~\cite{Beneke:2013jia}
\begin{equation}
 V_{1/m^2}(\mathbf{p},\mathbf{p}') = -\frac{4\pi\alpha_s C_F}{\mathbf{q}^2}\left[\mathcal{V}_{1/m^2}\frac{\mathbf{q}^2}{m_t^2}+\mathcal{V}_{p}\frac{\mathbf{p}^2+\mathbf{p}'^2}{2m_t^2}\right],
\end{equation}
where $\mathbf{q}=\mathbf{p}-\mathbf{p}'$, the term proportional to
$\mathcal{V}_{1/m^2}$ vanishes for the P-wave due to asymmetry under
the integration over the spatial momentum components and
$\mathcal{V}_{p}=1+\mathcal{O}(\alpha_s)$. Our result for the
P-wave Green function expanded in $\alpha_s$ and $(1-z)$ reads
\begin{eqnarray}
 G_P(1-z) & \mathop{\asymp}\limits^{z\to1} & \frac{m_t^4}{4\pi}\Bigg\{ \left[(1-z)^{3/2} - \frac12 (1-z)^{5/2} + \mathcal{O}\left((1-z)^{7/2}\right)\right] \nonumber\\
        &        & + \alpha_s C_F \left[\frac12 (1-z)\ln(1-z) - (1-z)^2 \ln(1-z) + \mathcal{O}\left((1-z)^3\right)\right] \nonumber\\
        &        & + \alpha_s^2 C_F \Big[-C_F\,\frac{3+\pi^2}{12}\sqrt{1-z} \nonumber\\ 
        &        & \hspace{1.6cm} - \frac{1-z}{16\pi}\left(\beta_0\ln^2(1-z)-2(a_1+2\beta_0)\ln(1-z)\right)\nonumber\\
        &        & \hspace{1.6cm} +\, \mathcal{O}\left((1-z)^{3/2}\right)\Big]  + \mathcal{O}(\alpha_s^3)\Bigg\},
\end{eqnarray}
where
\begin{equation}
 \beta_0 = \frac{11}{3}C_A-\frac{4}{3}T_F n_l, \hspace{1cm} a_1 = \frac{31}{9}C_A-\frac{20}{9}T_F n_l,
\end{equation}
and $\asymp$ again indicates that terms analytic in $(1-z)$
have been dropped.

\section{Results for the $gg\to HH$ form factors near threshold\label{sec:ggHH_Results}}

\begin{figure}[t!]
 \centering
\includegraphics[width=0.7\textwidth]{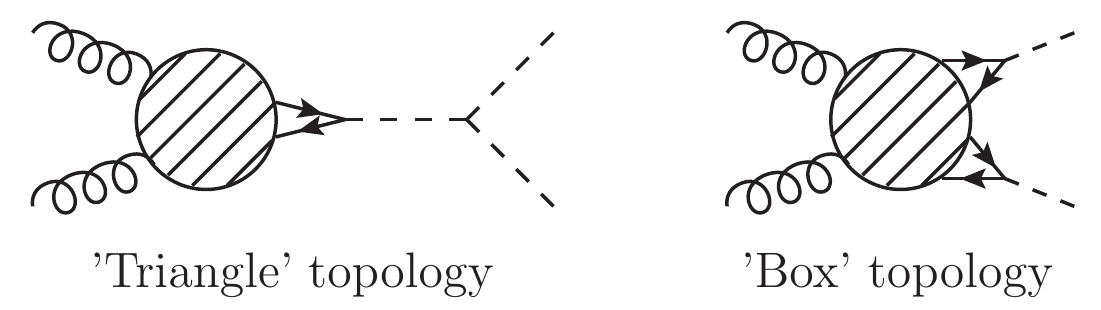}
\caption{The Feynman diagrams are divided into 'triangle' and 'box'
topologies depending on whether they contain an intermediate
s-channel Higgs boson or not. \label{fig:tri_vs_box}}
 \end{figure}

We give the results for the threshold expansion of the $gg\to HH$ form
factors up to three-loop order. The expansion of the form factors
$F_1$ and $F_2$ in the strong coupling constant takes the form
\begin{equation}
 F_i = F_i^{1l} + \frac{\alpha_s}{\pi}\,\left[F_i^{2l} + F_i^{2\triangle}\right]\, + \left(\frac{\alpha_s}{\pi}\right)^2\,\widetilde{F}_i^{3l} + \dots, \hspace{1.5cm}i=1,2.
 \label{eq:FiAlsExpansion}
\end{equation}
At the two-loop level the contributions $F_i^{2\triangle}$ that involve
two top-quark loops are known exactly~\cite{Degrassi:2016vss} and have
therefore been separated in~\eqref{eq:FiAlsExpansion}. They will not be
considered further because their threshold expansion does not contain
any non-analytic terms. The form factor $F_1$ is further decomposed
into a 'triangle' and 'box' contribution
\begin{equation}
 F_1^{il} = \frac{3 m_H^2}{\hat{s} - m_H^2}\,F_\triangle^{il} + F_\square^{il}, \hspace{1.5cm}i=1,2.  \label{eq:f1}
\end{equation}
as indicated in Figure~\ref{fig:tri_vs_box}. The contributions of the
'triangle' diagrams to the form factor $F_2$ vanish.
As discussed in Section~\ref{sec:method} we make massless cuts explicit
\begin{eqnarray}
\label{eq:F2l_splitting}
 F_i^{2l} & = & C_F F_{i,C_F}^{2l} + C_A \left(F_{i,C_A}^{2l}+F_{i,C_A,\ln}^{2l}\ln(-4z)\right), \\
 \widetilde{F}_i^{3l} & = & F_i^{3l} + F_{i,\ln}^{3l}\ln(-4z) + F_{i,\ln^2}^{3l}\ln(-4z)^2, \hspace{2.5cm}i=\triangle,\square,2,
\label{eq:F3l_splitting}
\end{eqnarray}
such that all the $F$'s on the right-hand side of \eqref{eq:F2l_splitting}
and \eqref{eq:F3l_splitting} are analytic in $z$ except for a branch cut
for real $z\geq1$.

Like the previous works~\cite{Spira:1995rr,Harlander:2005rq,Degrassi:2016vss}
we use the following $\overline{\text{MS}}$ scheme convention
\begin{equation}
 \int d^4l \to \frac{\Gamma(1-\epsilon)}{(4\pi)^\epsilon}\mu_R^{2\epsilon}\int d^dl =
 \left[1 + \frac{\pi^2}{12} \epsilon^2+\dots\right]\frac{e^{\epsilon\gamma_E}}{(4\pi)^\epsilon}\mu_R^{2\epsilon}\int d^dl
\end{equation}
in our calculation.
The renormalized form factors still contain IR divergences which cancel
with contributions involving unresolved real radiation that are not considered
here. We use subtractions of a minimal type as
Refs.~\cite{Spira:1995rr,Harlander:2005rq,Degrassi:2016vss}~\footnote{
In spite of some notational differences,
our convention is identical to~\cite{Degrassi:2016vss}. There is an exact
cancellation between the $\beta_0/\epsilon$ contribution in~\eqref{eq:IRsubtraction2l}
and the charge and gluon field renormalization terms. This has been exploited
in~\cite{Degrassi:2016vss} where both effects are not written explicitly.
Furthermore, the sign in the factor $(-\hat{s}-i0)^{-\epsilon}$ has been ignored
in~\cite{Degrassi:2016vss} because the induced imaginary part is not relevant
within the LME at the considered order in the strong coupling constant.}
\begin{eqnarray}
\label{eq:IRsubtraction2l}
 F_{i}^{2l} & = & F_{i,\text{virt+ct.}}^{2l} + \left[\frac{C_A}{2\epsilon^2}\,\left(\frac{\mu_R^2}{-\hat{s}-i0}\right)^\epsilon+\frac{\beta_0}{4\epsilon}\right]F_{i}^{1l},\\
 \widetilde{F}_{i}^{3l} & \mathop{\asymp}\limits^{z\to1} &
 \widetilde{F}_{i,\text{virt+ct.}}^{3l} + \left[\frac{C_A}{2\epsilon^2}\,\left(\frac{\mu_R^2}{-\hat{s}-i0}\right)^\epsilon+\frac{\beta_0}{4\epsilon}\right]F_{i}^{2l} + \mathcal{O}((1-z)^{3/2}).
\end{eqnarray}
The full form of the subtraction term at NNLO is known~\cite{Grigo:2015dia,Catani:1998bh}
and includes a contribution proportional to $F_{i}^{1l}$ which has been omitted here
because it only affects the three-loop results beyond the considered order in the
threshold expansion.

Our results for the triangle form factor at one and two loops are given
in~(\ref{eq:triangle_threshold1}-\ref{eq:triangle_threshold4}).
At the one-loop order we determine the remaining form factors up to nrNNLO
in the threshold expansion
\begin{eqnarray}
 F_\square^{1l} & \mathop{\asymp}\limits^{z\to1} & - \frac{2\pi(5-8r_H)}{3(1-2r_H)^2} (1-z)^{3/2}  - \frac{\pi}{15 (1-2 r_H)^4} (1-z)^{5/2} \nonumber \\
 & & \times \left[147-16 r_{p_T}-r_H (836-64 r_{p_T})+4 r_H^2 (409-16 r_{p_T})-1056 r_H^3\right] \nonumber\\
 & & + \mathcal{O}\left((1-z)^{7/2}\right), \label{eq:thres_F1LO}\\
 F_2^{1l} & \mathop{\asymp}\limits^{z\to1} & - \frac{8 \pi\, r_{p_T}}{3 (1-2 r_H)^2} (1-z)^{3/2}
 - \frac{4 \pi\, r_{p_T} \left(29-100 r_H+108 r_H^2\right)}{15 (1-2 r_H)^4} (1-z)^{5/2} \nonumber \\
 & & + \mathcal{O}\left((1-z)^{7/2}\right). \label{eq:thres_F2LO}
\end{eqnarray}
At two-loop order the threshold expansion up to nrNNLO takes the form
\begin{eqnarray}
 F_{\square,C_F}^{2l} & \mathop{\asymp}\limits^{z\to1} & - \frac{\pi ^2 (5-8 r_H)}{3 (1-2 r_H)^2} (1-z) \ln(1-z) + \Bigg[\frac{\pi}{12 (1 - 2r_H)^2(1 - 4 r_H)} \nonumber \\
 & & \Bigg(64 - 3 \pi^2 - 32 \ln(2) - \left[32 + 12 \pi^2 - 192 \ln(2)\right] r_{p_T}  \nonumber\\
 & & - \left[416 - 12 \pi^2 - 256 \ln(2) - \left(128 + 48 \pi^2 - 768 \ln(2)\right) r_{p_T}\right] r_H \nonumber\\
 & & + 16 \left[41 - 32 \ln(2)\right] r_H^2 - 128 r_H^3\Bigg)  \nonumber\\
 & & + \frac{8 \pi \left(1 - 9 r_H + 20 r_H^2 + 12 r_H^3 - 40 r_H^4\right)}{3 (1 - 2r_H)^2(1 - 4 r_H)^2}\ln(2 - 4 r_H) \nonumber\\
 & & + \frac{4\pi(3 - 10 r_H + 16 r_H^2 - 12 r_H^3)}{3 (1-2 r_H)^3}\sqrt{\frac{1 - r_H}{r_H}}  \arctan\left(\frac{2 \sqrt{r_H(1 - r_H)}}{1 - 2 r_H}\right) \nonumber\\
 & & + \frac{4 \pi (2 - 7 r_H + 2 r_H^2)}{3 (1 - 2 r_H)}C_0(1,4r_H,-1+4r_H;0,1,1)\Bigg] (1-z)^{3/2} - \frac{4 \pi ^2 }{15 (1-2 r_H)^4} \nonumber\\
 & & \times \left(9-2 r_{p_T}-4r_H (13-2 r_{p_T})+r_H^2 (107-8 r_{p_T})-72 r_H^3\right) (1-z)^2 \ln(1-z) \nonumber\\
 & & + \mathcal{O}\left((1-z)^{5/2}\right),\\
 F_{\square,C_A}^{2l} & \mathop{\asymp}\limits^{z\to1} & -\frac{\pi }{6 (1-2 r_H)^2}\Big[2 - 2\pi^2 - 4 \ln(2) + \left[2 \pi^2 + 16 \ln(2)\right]r_H \nonumber\\
 & & + \left[8 - 3 \pi^2 + 24 \ln(2)\right]r_{p_T}\Big] (1-z)^{3/2} + \mathcal{O}\left((1-z)^{5/2}\right),\\
 F_{\square,C_A,\ln}^{2l} & \mathop{\asymp}\limits^{z\to1} & \mathcal{O}\left((1-z)^{5/2}\right),\\
 F_{2,C_F}^{2l} & \mathop{\asymp}\limits^{z\to1} & - \frac{4\pi ^2 r_{p_T}}{3 (1-2 r_H)^2} (1-z) \ln(1-z) \nonumber \\
 & & + \Bigg[ \frac{4\, \pi\, r_{p_T}}{3 (1 - 2 r_H)^2 (1 - 4 r_H)^2} \Big(13 - 76 r_H + 116 r_H^2 - 16 r_H^3\Big) \nonumber \\
 & & - \frac{64\, \pi\, r_{p_T}(2 - 15 r_H + 37 r_H^2 - 36 r_H^3 + 16 r_H^4)}{3 (1 - 2 r_H)^2 (1 - 4 r_H)^3} \ln(2 - 4 r_H) \nonumber\\
 & & - \frac{16\, \pi\,  r_{p_T} r_H (14 - 67 r_H + 92 r_H^2 - 36 r_H^3)}{3 (1 - 2 r_H)^3 (1 - 4 r_H)^2}\sqrt{\frac{1 - r_H}{r_H}}  \arctan\left(\frac{2 \sqrt{r_H(1 - r_H)}}{1 - 2 r_H}\right) \nonumber\\
 & & - \frac{16\, \pi\, r_{p_T}(5 - 12 r_H + 6 r_H^2 + 4 r_H^3)}{3 (1 - 2 r_H) (1 - 4 r_H)^2} C_0(1,4r_H,-1+4r_H;0,1,1) \Bigg] (1-z)^{3/2}  \nonumber\\
 & & - \frac{4 \pi ^2 r_{p_T}\left(7-20 r_H+24 r_H^2\right)}{15 (1-2 r_H)^4} (1-z)^2 \ln(1-z) + \mathcal{O}\left((1-z)^{5/2}\right),\\
 F_{2,C_A}^{2l} & \mathop{\asymp}\limits^{z\to1} & -\frac{2\,\pi\, r_{p_T}}{9 (1-2 r_H)^2}\left(2 - 3 \pi^2 + 10 \ln(2)\right) (1-z)^{3/2}   + \mathcal{O}\left((1-z)^{5/2}\right),\\
 F_{2,C_A,\ln}^{2l} & \mathop{\asymp}\limits^{z\to1} & -\frac{22\,\pi\, r_{p_T}}{9 (1-2 r_H)^2}(1-z)^{3/2}  + \mathcal{O}\left((1-z)^{5/2}\right).
\end{eqnarray}
The three-loop form factors are determined at nrNLO in the threshold expansion
\begin{eqnarray}
 F_{\triangle}^{3l} & \mathop{\asymp}\limits^{z\to1} & -\frac{\pi ^3}{6} \left(3+\pi ^2\right) C_F^2 \sqrt{1-z}
     + \frac{\pi^2C_F(1-z)}{8}\Bigg\{ - \beta_0 \ln ^2(1-z) \\
 & & + \left[2 a_1+4 \beta_0-\left(\pi^2-\frac{4}{3}\right) C_A-\left(\frac{40}{3}-\pi^2\right) C_F\right]\ln (1-z)
 \Bigg\} + \mathcal{O}\left((1-z)^{3/2}\right), \nonumber\\
 F_{\square}^{3l}   & \mathop{\asymp}\limits^{z\to1} & \frac{\pi ^3 \left(3+\pi ^2\right) C_F^2 (5-8 r_H) \sqrt{1-z}}{18 (1-2 r_H)^2}
     + \frac{\pi^2C_F(1-z)}{24(1-2 r_H)^2}\Bigg\{ \beta_0 (5-8 r_H) \ln ^2(1-z) \nonumber\\
 & & + \Bigg[-2 (a_1+2\beta_0) (5-8 r_H) - 4 C_A \Bigg(1-2\ln(2)-\pi^2 + (8 \ln(2)+\pi^2) r_H \nonumber\\
 & & + \left(4+12\ln(2)-\frac{3\pi^2}{2}\right) r_{p_T}\Bigg)
      + C_F\Bigg[\frac{-1}{1 - 4 r_H}  + 65 - 32 \ln(2)-3 \pi ^2  \nonumber\\
 & & - (32-192 \ln(2)+12\pi^2) r_{p_T} - 4 (39-32\ln(2)) r_H + 32 r_H^2 \nonumber\\
 & & + \frac{32 \left(1-9 r_H+20 r_H^2+12 r_H^3-40 r_H^4\right)}{(1-4 r_H)^2} \ln(2 - 4 r_H) \nonumber\\
 & & + \frac{16 \left(3-10 r_H+16 r_H^2-12 r_H^3\right)}{1-2 r_H}\sqrt{\frac{1 - r_H}{r_H}}  \arctan\left(\frac{2 \sqrt{r_H(1 - r_H)}}{1 - 2 r_H}\right) \nonumber\\
 & & + 16 (1-2 r_H) \left(2-7 r_H+2 r_H^2\right) C_0(1,4r_H,-1+4r_H;0,1,1)\Bigg]\Bigg]\ln(1-z)\Bigg\} \nonumber\\
 & & + \mathcal{O}\left((1-z)^{3/2}\right),\\
 F_{2}^{3l}         & \mathop{\asymp}\limits^{z\to1} & \frac{2 \pi ^3 \left(3+\pi ^2\right) C_F^2 r_{p_T} \sqrt{1-z}}{9 (1-2 r_H)^2}
     + \frac{\pi^2C_F r_{p_T}(1-z)}{6(1-2 r_H)^2}\Bigg\{ \beta_0 \ln ^2(1-z) \nonumber\\
 & & + \Bigg[-2(a_1+2\beta_0) - \frac{2}{3} C_A \left(2-3 \pi ^2+10 \ln(2)\right) + \frac{4C_F}{(1-4r_H)^2}\Bigg[ 13 \nonumber\\
 & & - 76 r_H+116 r_H^2-16 r_H^3 - \frac{16 \left(2-15 r_H+37 r_H^2-36 r_H^3+16 r_H^4\right)}{1-4 r_H} \ln(2 - 4 r_H)\nonumber\\
 & & - \frac{4 r_H \left(14-67 r_H+92 r_H^2-36 r_H^3\right)}{1-2 r_H}\sqrt{\frac{1 - r_H}{r_H}}  \arctan\left(\frac{2 \sqrt{r_H(1 - r_H)}}{1 - 2 r_H}\right) \nonumber\\
 & & - 4 (1-2 r_H) \left(5-12 r_H+6 r_H^2+4 r_H^3\right) C_0(1,4r_H,-1+4r_H;0,1,1) \Bigg]\Bigg]\ln(1-z) \Bigg\} \nonumber\\
 & & + \mathcal{O}\left((1-z)^{3/2}\right),\\
 F_{2,\ln}^{3l}     & \mathop{\asymp}\limits^{z\to1} & -\frac{11 \pi^2 C_F C_A r_{p_T} (1-z)\ln(1-z)}{9 (1-2 r_H)^2} + \mathcal{O}\left((1-z)^{3/2}\right).
\end{eqnarray}
The logarithmic coefficients of~\eqref{eq:F3l_splitting} that are not
written explicit above vanish up to and including the order $(1-z)$.
The scalar triangle integral appearing above is given by
\begin{equation}
 C_0(1,4r_H,-1+4r_H;0,1,1) = \int\frac{d^dl}{i\pi^{d/2}}\frac{m_t^2}{[l^2][(l+q)^2-m_t^2][(l+q-p_H)^2-m_t^2]},
\end{equation}
with $q^2=m_t^2$, $p_H^2=m_H^2$ and $q\cdot p_H=m_t^2$. All boxes that
appear in the hard matching computation can be reduced to at most
triangles by partial fractioning since only three of the propagators
in each box are linearly independent at threshold.

\end{document}